\documentclass[%
reprint, 
10pt,
superscriptaddress,
amsmath,amssymb, aps,
pra,
]{revtex4-2}
\usepackage{amsmath}
\usepackage{xcolor}
\usepackage{graphicx}
\usepackage{dcolumn}
\usepackage{bm}
\usepackage[english]{babel}
\usepackage{url}

\usepackage{changes}

\usepackage[normalem]{ulem}
\usepackage{graphicx} 
\usepackage{natbib} 
\usepackage{hyperref}
\usepackage{ifthen} 
\usepackage{tikz}
\usetikzlibrary{decorations.pathmorphing,decorations.markings}
\usepackage{makeidx}
\usepackage{amssymb} 
\usepackage{braket}
\usepackage{mathdots}

\begin{document}

\preprint{APS/123-QED}

\title{Spectroscopy and Coherent Control of Two-Level System Defect Ensembles Using a Broadband 3D Waveguide}

\author{Qianxu Wang}
\thanks{These authors contributed equally to this work.}
\affiliation{\mbox{Department of Physics and Astronomy, Dartmouth College, 6127 Wilder Laboratory, Hanover, New Hampshire 03755, USA}}
\affiliation{\mbox{Thayer School of Engineering, Dartmouth College, 15 Thayer Drive, Hanover, New Hampshire 03755, USA}}

\author{Juan S. Salcedo-Gallo}
\thanks{These authors contributed equally to this work.}
\affiliation{\mbox{Thayer School of Engineering, Dartmouth College, 15 Thayer Drive, Hanover, New Hampshire 03755, USA}}

\author{Salil Bedkihal}
\thanks{These authors contributed equally to this work.}
\affiliation{\mbox{Thayer School of Engineering, Dartmouth College, 15 Thayer Drive, Hanover, New Hampshire 03755, USA}}

\author{Tian Xia}
\affiliation{\mbox{Thayer School of Engineering, Dartmouth College, 15 Thayer Drive, Hanover, New Hampshire 03755, USA}}

\author{Maciej W. Olszewski}
\affiliation{\mbox{Department of Physics, Cornell University, Ithaca, NY 14853, USA}}

\author{Valla Fatemi}
\affiliation{\mbox{School of Applied and Engineering Physics, Cornell University, Ithaca, NY 14853, USA}}

\author{Mattias Fitzpatrick}
\email{mattias.w.fitzpatrick@dartmouth.edu}
\affiliation{\mbox{Thayer School of Engineering, Dartmouth College, 15 Thayer Drive, Hanover, New Hampshire 03755, USA}}
\affiliation{\mbox{Department of Physics and Astronomy, Dartmouth College, 6127 Wilder Laboratory, Hanover, New Hampshire 03755, USA}}

\begin{abstract}
Defects in solid-state materials play a central role in determining coherence, stability, and performance in quantum technologies. Although narrowband techniques can probe specific resonances with high precision, a broadband spectroscopic approach captures the full spectrum of defect properties and dynamics. Two-level system (TLS) defects in amorphous dielectrics are a particularly important example because they are major sources of decoherence and energy loss in superconducting quantum devices. However, accessing and characterizing their collective dynamics remains far more challenging than probing individual TLS defects. Building on our previously developed Broadband Cryogenic Transient Dielectric Spectroscopy (BCTDS) technique, we study the coherent control and time-resolved dynamics of TLS defect ensembles over a wide frequency range of 3-5 GHz without requiring full device fabrication. Unlike conventional approaches limited to narrow spectral windows and fully fabricated devices, BCTDS enables modular, device-independent measurements that reveal quantum interference effects, memory-dependent dynamics, and dressed-state evolution within the TLS defect bath. The spectral response reveals distinct V-shaped structures corresponding to the bare eigenmode frequencies. Using these features, we extract a TLS defect spectral density of $84$ GHz$^{-1}$ for a silicon sample, across a 4.1–4.6 GHz span. Furthermore, we systematically investigate amplitude- and phase-controlled interference fringes for multiple temperatures and inter-pulse delays, providing direct evidence of coherent dynamics and control. A driven minimal spin model with dipole–dipole interactions that qualitatively capture the observed behavior is presented. Our results establish BCTDS as a versatile platform for broadband defect spectroscopy, offering new capabilities for diagnosing and mitigating sources of decoherence, engineering many-body dynamics, and exploring non-equilibrium phenomena in disordered quantum systems.
\end{abstract}

\date{\today}
\maketitle

\begin{figure}[hpbt!]
\begin{centering}
\includegraphics[width=0.42\textwidth]{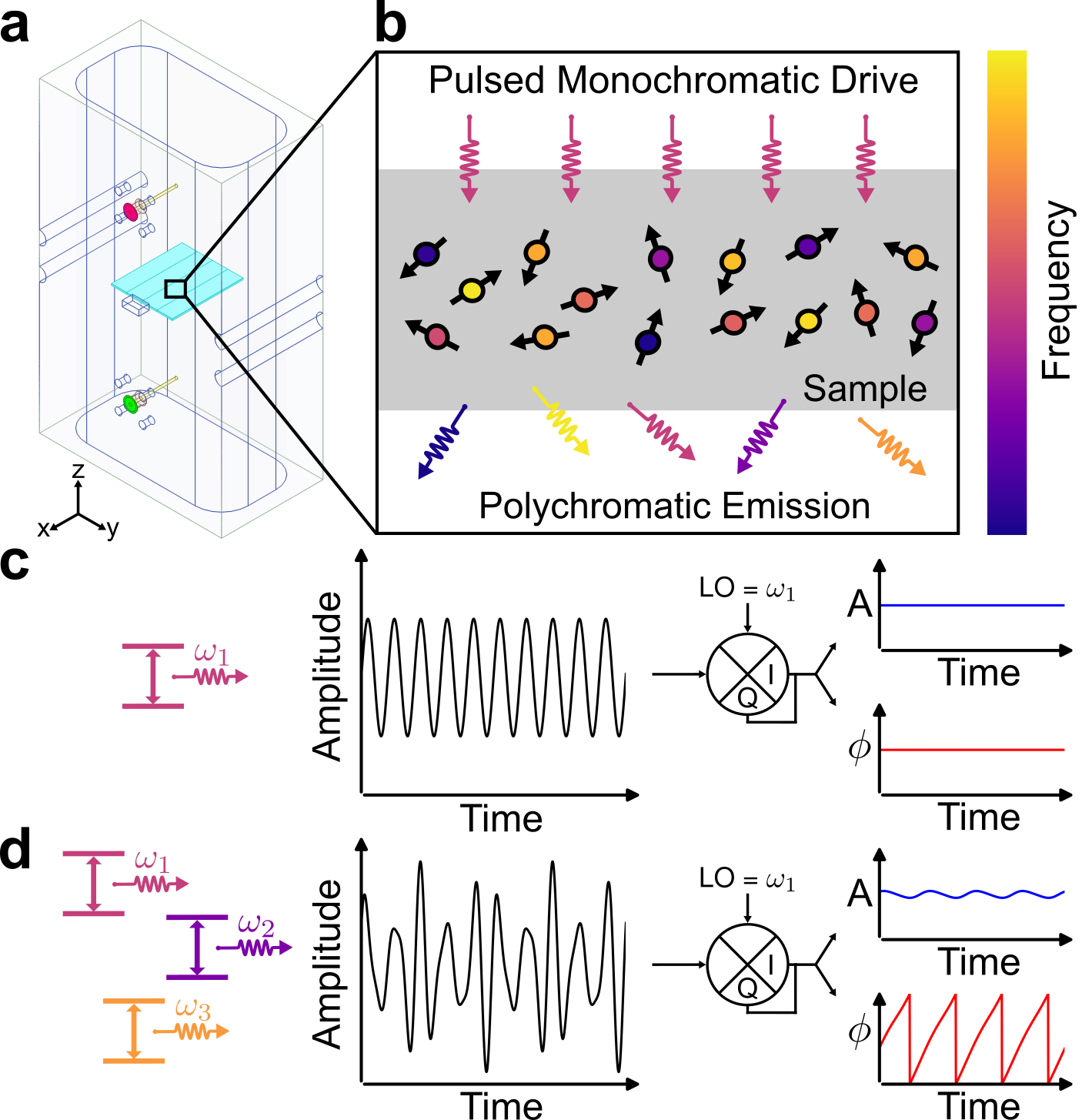}
\end{centering}
\caption{\label{fig:overview} 
BCTDS measurement setup and homodyne detection of polychromatic emission. \textbf{a,} Schematic of the broadband waveguide setup, with samples (cyan) mounted in the middle of the waveguide. 
\textbf{b,} Conceptual illustration of TLS defects in a sample, represented as a spin bath containing various defects at their respective resonance frequencies (color). Under a pulsed monochromatic drive, these defects become dressed and subsequently emit polychromatic radiation with frequency components described by Floquet-like sidebands and quasi-energies. 
\textbf{c,} Simple example of homodyne detection for a single emitter at angular frequency $\omega_1$. Mixing the signal with a local oscillator (LO) at $\omega_1$ removes the fast carrier oscillation, producing the in-phase ($I$) and quadrature ($Q$) signals from which we extract the amplitude and phase of the homodyne signal. In the case of monochromatic emission, the amplitude and phase remain constant when the local oscillator frequency matches the emitted signal.
\textbf{d,} Homodyne detection of a superposed signal from three independent emitters at angular frequencies $\omega_1$, $\omega_2$, and $\omega_3$. Homodyne detection at  $\omega_1$ results in an oscillating amplitude, and the linear phase evolution results in a characteristic sawtooth pattern due to the phase wrapping. The amplitude and phase carry information about all of the emitted frequencies that we can reconstruct through a Fast Fourier Transform (FFT).} \end{figure}

\section{\label{sec:intro} Introduction}
Defects in solid-state materials are both sources of decoherence and building blocks for quantum technologies, as certain vacancy–impurity complexes and dopants host long-lived, controllable spin states. These spin-active defects are studied in materials like diamond, silicon carbide, and silicon, enabling quantum computing, communication, and sensing \cite{RevModPhys.89.035002, Liu2019, Angerer2018, Wolfowicz2021, Fuchs2011, Yao2012, Kraus2014, PhysRevLett.125.107702, Esat2024, Choi2017, PhysRevApplied.20.L031001, PhysRevB.95.081405}.
In contrast, two-level system defects (TLS) within bulk materials, amorphous dielectrics, and interfaces are key sources of decoherence, spectral diffusion, energy and dielectric loss in superconducting circuits, albeit their microscopic origin and collective behavior remain unclear \cite{Siddiqi2021, PhysRevB.92.174201, PhysRevApplied.19.034064, Hegedus2025}. TLS detection typically relies on qubits or resonators, which probe only narrow frequency windows after device fabrication, limiting diagnostics during the intermediate stages of fabrication \cite{Burnett2014, PhysRevApplied.12.014012, deGraaf2020TLS, PhysRevApplied.19.024006, Müller_2019, PhysRevLett.111.065901, Lisenfeld2015, Lisenfeld2019, Biznarova2024}.

In the standard tunneling model, defects are treated as TLS ensembles with nearly degenerate states separated by a barrier, where low-temperature dynamics are dominated by quantum tunneling \cite{Anderson1972_1, Phillips1972, PhysRevB.88.174202}. High densities of parasitic TLS defects are believed to reside in tunnel barriers of Al/AlOx/Al Josephson junctions, native oxides on aluminum and niobium, and thin-film dielectrics \cite{Siddiqi2021}. The coherent signatures of these defects were first observed in phase qubits as avoided level crossings \cite{PhysRevLett.105.177001}, with tunability via mechanical strain and electric fields \cite{Lisenfeld2015}. Resonant coupling between qubits and TLS defects enables coherent energy exchange, degrading qubit performance, while repeated measurements of $T_1$ reveal strong temporal fluctuations linked to ergodic-like spectral diffusion of TLS ensembles \cite{Carroll2022, PhysRevB.104.094106}. These analyses suggest that the density of low-frequency TLS defects is roughly three orders of magnitude higher than that of high-frequency ones. Moreover, a two-time relaxometry technique has been recently introduced as a novel method for simultaneously and efficiently probing both the qubit and the TLS bath \cite{zhuang2025}. This approach has been applied to high-coherence fluxonium qubits over a frequency range of $0.1–0.4$ GHz, revealing a discrete spectrum of low-frequency TLS defects. This spectrum is consistent with a random distribution of TLS defects within the aluminum oxide tunnel barrier of the Josephson junction chain of the fluxonium with 0.4 GHz$^{-1}$$\mu$m$^{-2}$ as well as an average electric dipole moment of 6 Debye, comparable to those observed in previous TLS studies at much higher frequencies.

While the above techniques solely rely on narrow-band probes, it is important to understand the spectral signatures of TLS defect dynamics over a broadband. For instance, parametric amplification is crucial for high-fidelity qubit readout in cQED, with Josephson traveling wave parametric amplifiers (JTWPAs) offering high gain and multi-GHz bandwidth \cite{Macklin2015}. Recent work reported microsecond dielectric echoes in JTWPAs under strong drive, attributed to ensembles of excited defects \cite{boselli2025}. These long-lived echoes can impair fast readout, underscoring the need to characterize and mitigate collective TLS defect dynamics.

We recently introduced \emph{Broadband Cryogenic Transient Dielectric Spectroscopy} (BCTDS) as a technique to probe TLS defect dynamics over a wide frequency range in a single experiment, using a 3D microwave waveguide capable of measuring various sample configurations \cite{BCTDS}. By analyzing two-time intensity correlation functions obtained by homodyne measurements, BCTDS allows for direct extraction of the complex susceptibility and dielectric response of the materials~\cite{BCTDS}. This method enables modular, component-level characterization throughout fabrication and uncovers spectral features of collective TLS ensemble dynamics, including sideband signatures of off-resonant transitions that remain inaccessible to conventional narrowband qubit- or resonator-based probes. In addition to revealing TLS defect ensemble dynamics, the broadband 3D waveguide underlying BCTDS enables controlled studies of non-equilibrium behavior in disordered many-body systems. In a representative case, a sample with amorphous structures that hosts an ensemble of TLS defects is driven out of equilibrium by a pulsed field that implements a \emph{quench}. The subsequent broadband response captures transient ring-down with spectral signatures such as sidebands and beating, enabling analysis of relaxation dynamics and collective behavior~\cite{BCTDS}. The flexibility of the broadband 3D waveguide also supports measurements across diverse defect species, providing a modular approach.

In this work, we experimentally demonstrate coherent control of TLS defect ensembles in various samples using BCTDS. The technique probes quantum interference in driven ensembles and enables the extraction of their bare eigenmode frequencies. Using microwave pulses, we reveal memory effects in the ensemble dynamics and observe beating from sideband interference. By adjusting pulse amplitude and the relative phase and spacing between excitation pulses, we control the interference pattern, demonstrating manipulation of the TLS ensemble and providing a way to explore Floquet-engineered dynamics in many-body quantum systems~\cite{PhysRevApplied.18.064023}. This capability is crucial for advancing our understanding of disordered quantum systems and for enabling scalable architectures in quantum technologies.

\section{\label{sec:setup} 
Experimental Setup}

\subsection{Apparatus}
In this experiment, we build upon our previously developed BCTDS platform \cite{BCTDS} to enable more systematic information extraction and investigation of interference and memory effects through the coherent control of bath dynamics across various samples. To achieve broadband probing, we mount the sample between two WR-229 waveguide adaptors, which form a frequency passband from 3 to 6 GHz when enclosed. The waveguide geometry is designed to favor propagation of the dominant TE$_{10}$ mode and to minimize parasitic cavity-like resonances within this frequency range. Each half of the waveguide contains an SMA antenna, connected to SMA cables for the injection and collection of microwave photons in and out of the waveguide. The assembly (shown in Fig.\,\ref{fig:overview}\textbf{a}) is mounted at the mixing chamber stage of a Bluefors LD400 dilution refrigerator, cooling the sample to below 10 mK to suppress thermal broadening and preserve the coherence of TLS defect features. The setup is described in the Appendix of \cite{BCTDS}. For GHz tone generation and homodyne readout, we use an RFSoC (Radio Frequency System on Chip) 4×2 board, a Field-Programmable Gate Array (FPGA) device capable of on-board pulse synthesis and homodyne readout. This device enables us to synthesize pulses, excite the TLS defect bath, and collect the response during the post-pulse time period. This response is then amplified by a HEMT (High Electron Mobility Transistor) and several room-temperature amplifiers (Minicircuits CMA-83LN+).

\subsection{Operating Principle}
The conceptual basis for probing the transient dielectric response can be understood within the simple electrodynamics framework described below. When a dielectric sample placed in a broadband waveguide is driven by a coherent microwave tone of frequency $\omega$, the incident classical field $\mathbf{E}_{\mathrm{in}}(t) = \mathbf{E}_0 \cos (\omega t)$ induces a macroscopic polarization $\mathbf{P}(\mathbf{r}, t)$ within the sample. In the frequency domain, and assuming negligible spatial dispersion, the polarization can be approximated as

\begin{equation}
\begin{aligned}
    \mathbf{P}(\mathbf{r}, \omega) &\approx \varepsilon_0 \bm{\chi}(\mathbf{r}, \omega) \mathbf{E}_{\mathrm{loc}}(\mathbf{r, \omega}),
\end{aligned}
\end{equation}
where $\mathbf{E}_{\mathrm{loc}}(\omega) = L(\omega) \mathbf{E}_{\mathrm{in}}(\omega)$ is the local field. $L(\omega)$ is the local field correction.  Coarse-graining to a single dipole gives the effective dipole moment
\begin{equation} \label{eq:effective-dipole}
    \mathbf{p}(\omega) =  \int_V \mathbf{P}(\mathbf{r}, \omega) \, d^3r \approx \varepsilon_0 \bm{\chi}_{\mathrm{{eff}}}(\omega) \mathbf{E}_{\mathrm{loc}}(\omega),
\end{equation}
where $\bm{\chi}_{\mathrm{{eff}}}(\omega)$ is the effective susceptibility that describes the response of the sample. Assume from hereafter without loss of generality that the input field is purely along the $\hat{\mathbf{x}}$ direction. The dipole radiates back into the waveguide field, producing an output field
\begin{equation}
    {E}_{\mathrm{out}}(\omega) = t(\omega) {E}_{\mathrm{in}}(\omega) + i \bm{\alpha}(\omega) \mathbf{p}(\omega).
\end{equation}
Here $t(\omega)$ is the bare transmission of the waveguide within the dipole and $\bm{\alpha}(\omega)$ characterizes the coupling between the dipole and the measured output mode. Inserting the dipole relation in Eq. \ref{eq:effective-dipole} yields the effective transfer function
\begin{equation}
    T(\omega) = t(\omega) + i \varepsilon_0 \bm{\alpha}(\omega)  \bm{\chi}_{\mathrm{{eff}}}(\omega) L(\omega),
\end{equation}
such that 
\begin{equation}
    E_{\mathrm{out}}(\omega) = T(\omega) E_{\mathrm{in}}(\omega).
\end{equation}
Measurement of the amplitude and phase of the output field reveals the complex susceptibility \(\bm{\chi}_{\mathrm{eff}}(\omega)\), providing access to the dispersive and absorptive properties of the sample. 

In addition to the coherent response, the sample emits radiation resulting from its internally driven quantum dynamics. We model the input and output modes with the annihilation operators $\hat{a}_{\mathrm{in}}(t), \hat{a}_{\mathrm{out}}(t)$. A driven sample adds a field into the output mode:
\begin{equation}
\hat{a}_{\mathrm{out}}(t) = t^* \hat{a}_{\mathrm{in}}(t) + \beta^* \hat{s}(t),
\end{equation}
where $\hat{s}(t)$ represents the sample dipole fluctuations, and $t^*, \beta^*$ coupling coefficients. Fluctuations from the sample's driven dynamics contribute incoherent radiation determined by the two-time correlation function \(\langle \hat{s}^\dagger(t) \hat{s}(t+\tau) \rangle\), which influences the spectral density of the output field \cite{BCTDS}. Mixing the output field $\hat{a}_{\mathrm{out}}(t)$ with a local oscillator at frequency \(\omega_{\mathrm{LO}}\) and phase \(\phi_{\mathrm{LO}}\) produces baseband quadratures
\begin{equation}
\begin{aligned}
I(t) &\propto \mathrm{Re}\left[ e^{-i(\omega_{\mathrm{LO}} t + \phi_{\mathrm{LO}})}  \langle\hat{a}_{\mathrm{out}}(t)\rangle \right], \\
Q(t) &\propto \mathrm{Im}\left[ e^{-i(\omega_{\mathrm{LO}} t + \phi_{\mathrm{LO}})} \langle\hat{a}_{\mathrm{out}}(t)\rangle \right].
\end{aligned}
\end{equation}
Since $E_{\mathrm{out}}(t) \propto \langle\hat{a}_{\mathrm{out}}(t)\rangle$, the recorded signals are proportional to the output electrical field. Equivalently, we define the complex baseband signal
\begin{equation}
    z(t) = I(t) + iQ(t) \propto e^{-i(\omega_{\mathrm{LO}} t + \phi_{\mathrm{LO}})} E_\mathrm{out}(t),
\end{equation}
from which we reconstruct the instantaneous amplitude and phase
\begin{equation}
\begin{aligned}
    A(t) &= \sqrt{I^2(t) + Q^2(t)}\\
    \phi(t) &= \arg(I(t) + iQ(t)).
\end{aligned}
\end{equation}
With a single constant emission and a resonant LO, the baseband $I, Q$, and hence $A$ and  $\phi$ are time-independent (Fig.\,\ref{fig:overview}\textbf{c}). For a polychromatic emission from TLS ensembles with different resonances and sidebands, the homodyne readout
\begin{equation}
    z(t) \propto \sum_k c_k e^{-i(\omega_k - \omega_{\mathrm{LO}})t}
\end{equation}
produces amplitude envelopes with beating and phase trajectories with overall phase wrapping. The phase $\phi(t)$, which wraps around 0 and $2\pi$, shows a sawtooth pattern (Fig.\,\ref{fig:overview}\textbf{d}). Frequency information in amplitude and phase can be extracted using FFT, with amplitude FFT discussed previously \cite{BCTDS} and the phase FFT presented in Section \ref{sec:BCTDS and Phase V Structures}.

When the device is driven by a finite-duration microwave pulse for $0 <  t < t_\mathrm{off}$, the sample reaches a driven nonequilibrium state. After the drive is turned off at time $t = t_\mathrm{off}$, the system exhibits a \emph{transient ring-down} governed by its intrinsic eigenmodes with complex frequencies \(\tilde{\omega}_k = \omega_k - i \gamma_k\), where \(\gamma_k\) are the decay rates. The output field during ring-down is a sum of damped oscillations
\begin{equation}
E_{\mathrm{out}}(t) \propto \sum_k A_k e^{-i \omega_k (t - t_{\mathrm{off}})} e^{-\gamma_k (t - t_{\mathrm{off}})},
\end{equation}
which appear in the measured quadratures as damped oscillations at frequencies \(\omega_k - \omega_{\mathrm{LO}}\). In the \emph{quantum regime}, the post-pulse state includes non-linearities and coherent dynamics of the underlying two-level systems, which are fully captured by the measured quadratures and their correlation functions as reported in our earlier work \cite{BCTDS}.

\section{\label{sec:experimental_results} 
Experimental Results}

\subsection{\label{sec:BCTDS and Phase V Structures}
BCTDS and Phase V Structures
}
The BCTDS platform enables the collective excitation of TLS defect ensembles present in various materials and interfaces, allowing their complex dynamics to be probed through the transient response. To illustrate the rich information accessible with this approach, we consider an experiment in which a sample is interrogated using drive pulses of varying duration. We use a silicon wafer with (111) crystal orientation, purchased from WaferPro. The wafer is diced into rectangular chips with 27.5 × 5.5 mm dimensions and solvent cleaned to remove the dicing resist, with a natural silicon oxide layer subsequently reforming on the exposed surface. More details for this sample preparation are given in Appendix~\ref{app_sec:waferpro111_prep}. Previous characterizations indicate that surface oxides probably host a high density of defects, which contribute significantly to the pronounced responses observed under BCTDS (\cite{BCTDS}).

\begin{figure*}[htpb]
\includegraphics[width=\textwidth]{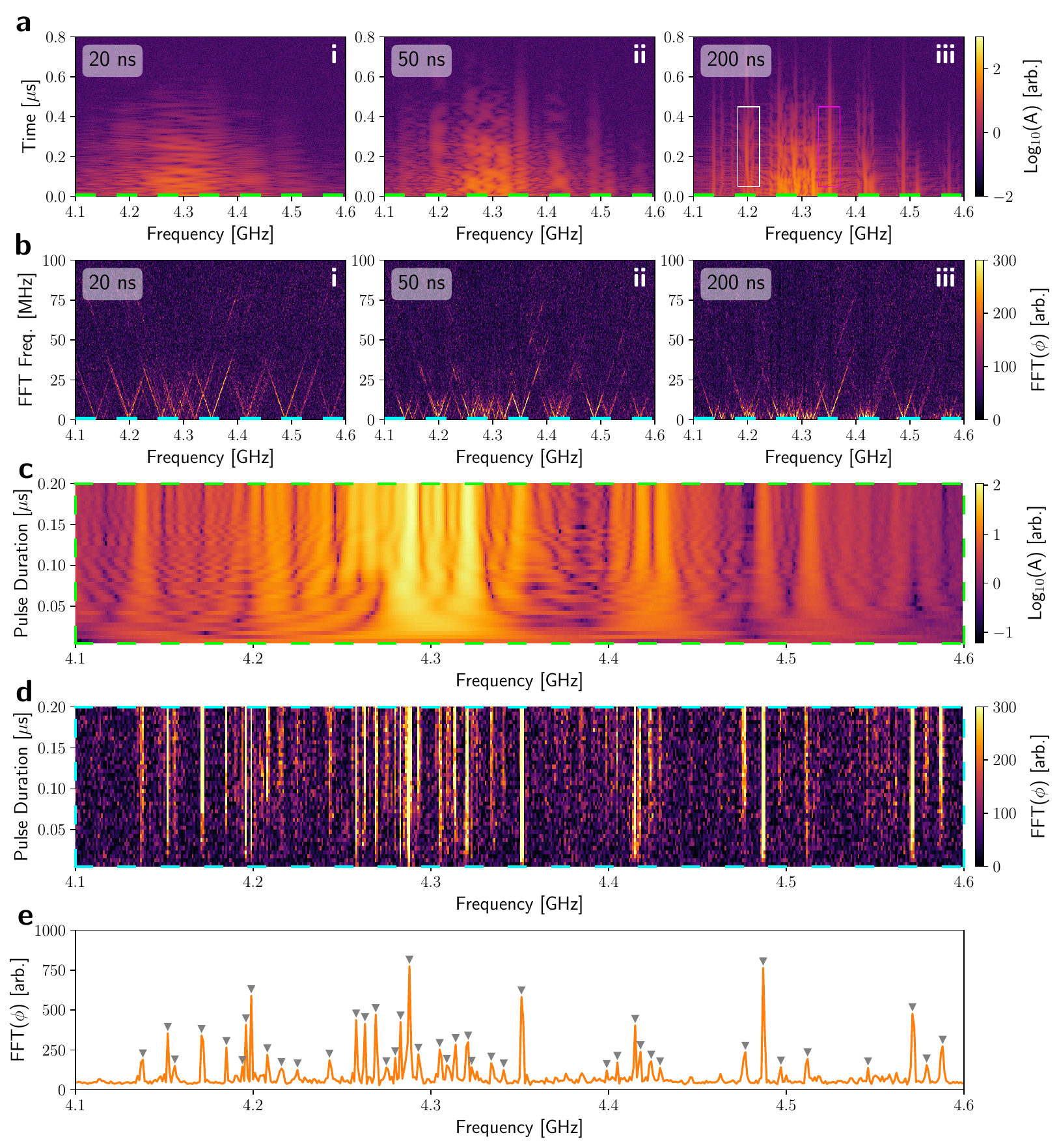}
\caption{BCTDS result for a silicon sample with silicon oxide under different drive durations. \textbf{a,} Logarithmic amplitude of the homodyne signal arising from the transient dielectric response.  \textbf{b,} FFT of the phase of the transient dielectric response, where V-shaped structures centered at the bare eigenfrequencies of TLS defect ensembles are clearly visible. Panels \textbf{a,b} are shown for three different pulse durations: 20 ns (\textbf{i}), 50 ns (\textbf{ii}), and 200 ns (\textbf{iii}). In \textbf{aiii}, we highlight a representative braiding pattern (white box) and single emission pattern (magenta box) resulting from emissions of different detunings. The drive amplitudes across the span are calibrated using a method described in \cite{BCTDS} Appendix E, to ensure constant amplitude driving across all frequencies. \textbf{c,} Zero-time transient response at different pulse durations. Horizontal slices (lime dashed lines) from the logarithmic amplitude plots (\textbf{a}) are taken at $t = t_{\text{off}} =0$ for a continuous sweep of pulse durations from 5 to 200 ns, showing sharpening interference patterns as the pulse duration increases. \textbf{d,} Base location of phase Vs at different pulse durations. Horizontal slices at 0 MHz (cyan dashed lines) from the phase FFT plots (\textbf{b}) are taken for the 5-200 ns duration sweep. Bright lines marking the eigenfrequencies of the driven system start to emerge, and their positions remain constant as the pulse duration increases. \textbf{e,} Zero-frequency slices of phase FFT magnitude, averaged over different pulse durations. Prominent peaks correspond to constant phase V base locations at various pulse durations, while incoherent fluctuations average out. Counting all peaks with magnitude $>$120 a.u. (marked by gray triangles) yields a quantitative estimate of the TLS defect spectral density of $84$ GHz$^{-1}$, across the 4.1-4.6 GHz span.}
\label{fig:phase_V}
\end{figure*}

The transient response of our silicon samples is illustrated in Fig.\,\ref{fig:phase_V}. The observed patterns are markedly different from prior control measurements on empty waveguides and clean sapphire samples at base temperature, which exhibit only sparse resonances \cite{BCTDS}. Furthermore, the patterns emerge exclusively at cryogenic temperatures and reconfigure to different frequency locations after thermal cycling (see Appendix~\ref{app_sec:temperature_dependence} and \ref{app_sec:thermal_cycle}), consistent with reports from previous literature \cite{Lisenfeld2015, PhysRevLett.105.177001}. We intentionally zoom into a frequency band of 4.1-4.6 GHz to resolve dense TLS defect activities. The sample is driven with a simple square pulse that excites ensembles of defects and dresses the system into Floquet states with associated sidebands. After the drive is switched off, the system relaxes by emitting energy near these sidebands, producing coherent ring-downs that exhibit beating from the interference of nearby resonances. The nature of this interference depends sensitively on the driving conditions. We explore this dependence by varying the drive duration from 5 to 200 ns and showing representative cases in Fig.\,\ref{fig:phase_V}\textbf{ai-iii}, corresponding to 20, 50, and 200 ns, respectively. As the pulse length increases, the spectral features become more refined, and sharper structures emerge in the transient response. This behavior is further demonstrated with zero-time slices, which are horizontal cuts taken immediately after the pulse is turned off, at $t = t_{\text{off}} =0$ (marked by lime dashed lines in Fig.\,\ref{fig:phase_V}\textbf{ai-iii}). For each duration in the 5–200 ns sweep, we extract the zero-time slice and stack the results to form the color plot in Fig.\,\ref{fig:phase_V}\textbf{c}. The pulse duration modifies how the sidebands and nearby resonances interfere with each other, giving rise to sharpening beating patterns between the dominant spectral features. This encouraging observation further suggests the possibility of coherently controlling the interference response through engineered pulse shapes, which we demonstrate in Section \ref{sec:coherent bath manipulation}.

In order to infer information about the driven system and the beating frequencies among different resonance components, we naturally turn to the FFT of the amplitude signal, as demonstrated in our previous BCTDS work \cite{BCTDS}. The resulting spectra exhibit characteristic V-shaped features that converge towards the dominant eigenfrequencies of the system. However, weaker resonances are often obscured by noise, limiting the amount of extractable information. A more subtle yet powerful method involves applying FFT to the phase of the homodyne signal (Fig.\,\ref{fig:phase_V}\textbf{bi-iii}). Just as amplitude oscillations encode interference between resonances, phase captures detuning between the homodyne signal and these various signal components. The phase evolution wraps around 2$\pi$, forming a “sawtooth” pattern that is especially easy to capture through frequency-domain analysis, as explained in the detuned multi-emitter example in Fig.\,\ref{fig:overview}\textbf{d}. This approach yields a much cleaner series of V-shaped spectral signatures, which convey rich information about the system, consistent with the ring-down patterns observed in the logarithmic amplitude spectrum (Fig.\,\ref{fig:phase_V}\textbf{a}).

The intensity of the V-shaped features reflects the orientation of the TLS defects. Brighter Vs likely indicates stronger alignment of the associated defect dipole with the polarized driving field. Correspondingly, at frequency locations with brighter V features, the logarithmic amplitude spectrum often exhibits longer and more pronounced ring-down patterns. In addition, we observe oscillations in the amplitude of the phase FFT, along the arm of V-shaped structures, that become faster as the drive duration increases, as clearly observed in Fig.\,\ref{fig:phase_V}\textbf{bi-iii}. 

As motivated in the later theory Section \ref{sec:floquet}, the positions of the V features map directly onto the eigenfrequencies of the driven system, offering essential insight into the frequency distribution of TLS defects and their behavior under the applied drive. In Fig.\,\ref{fig:phase_V}\textbf{aiii}, certain emissions appear closer in frequency and produce braided ring-down patterns, an example near 4.2 GHz is highlighted (white box). The corresponding phase FFT spectrum in Fig.\,\ref{fig:phase_V}\textbf{biii} shows multiple closely spaced V-shaped features around this frequency. By contrast, an isolated sharp emission appears near 4.35 GHz (magenta box), with a single, standalone V-shaped structure. If the TLS defects are weakly interacting and close in frequency, the small frequency difference between them can act as a detuning, giving rise to the observed braiding pattern. In contrast, if the TLS defects have identical frequencies, a single sharp emission is expected instead of a braiding pattern. To corroborate this interpretation, we simulate the dynamics of two TLS defects with a small frequency difference and with identical frequencies, as shown in Appendix \ref{app_sec:role_of_detuning}, Fig.\,\ref{fig:braiding_physics}. It is worth noting that the broadband nature of the waveguide is crucial to probing these intricate dynamics.

We further analyze the locations of V-shaped features by taking zero-frequency slices (marked by cyan dashed lines) of the phase FFT plots in Fig.\,\ref{fig:phase_V}\textbf{bi-iii}. Each slice produces a spectrogram whose peaks correspond to the base locations of the Vs. Repeating this procedure for a sweep of drive durations between 5 to 200 ns and stacking the resulting slices generates the color plot in Fig.\,\ref{fig:phase_V}\textbf{d}, which allows comparison of base V positions across pulse durations. The resulting bright vertical lines indicate that the locations of the Vs remain fixed as the pulse duration is varied, despite the substantial changes observed in the amplitude spectra (Fig.\,\ref{fig:phase_V}\textbf{ai–iii}). In Appendix \ref{app_sec:amp_sweep}, we also show the base V locations staying fixed while we sweep the amplitude of the drive tone and observe saturation behavior consistent with a nonlinear response. In some cases, weaker V features only emerge at longer pulse durations or stronger drive amplitudes, but their locations remain unchanged. These TLS defects likely have dipoles with weaker alignment to the driving field. Taken together, these results establish the important conclusion that the base positions of the Vs are independent of pulse parameters, consistent with their interpretation as the bare eigenfrequencies of the TLS defect. In contrast, thermal cycling often leads to reconfiguration of TLS defects, which can result in drastic changes in the V locations (Appendix \ref{app_sec:thermal_cycle}).

The invariance of the base V locations across different drive conditions allows us to determine the frequency distribution of TLS defects more reliably. By averaging zero-frequency slices taken at various pulse durations in Fig.\,\ref{fig:phase_V}\textbf{d}, we construct the spectrogram in Fig.\,\ref{fig:phase_V}\textbf{e}, where random noise is suppressed while the invariant peak locations persist. We identify 42 peak locations (marked by gray triangles) within the 4.1–4.6 GHz window, which yields an estimated defect density of $84$ GHz$^{-1}$, demonstrating a powerful and robust approach for rapid characterization and quantitative estimation of TLS defect spectral densities in different materials. 

\subsection{\label{sec:coherent bath manipulation}
Coherent Bath Manipulation of TLS Defects in Various Materials
}
\begin{figure}[ht]
\begin{centering}
\includegraphics[width=0.5\textwidth]{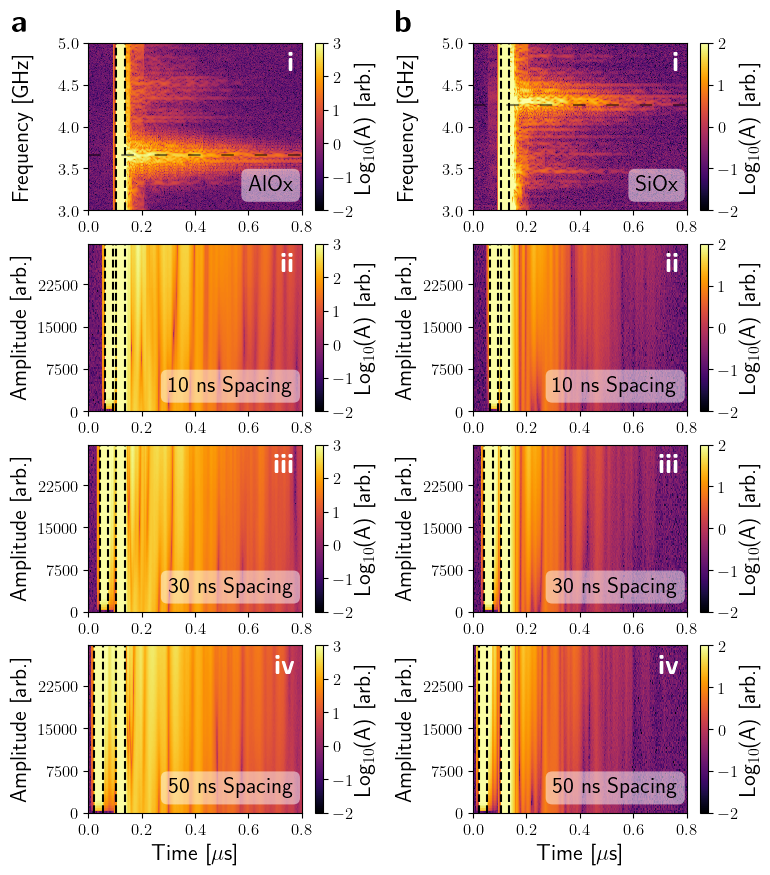}
\end{centering}
\caption{
Coherent control of TLS defect ensembles in different samples using pulse amplitude and spacing. \textbf{a,} Sapphire sample with a thin 2 nm AlOx deposit via Atomic Layer Deposition (ALD). \textbf{b,} Silicon samples with a native oxide layer (same sample as Fig.\,\ref{fig:phase_V}). Sub-panel \textbf{i} shows the single pulse BCTDS spectroscopy result for the two samples over a 3-5 GHz range. Sub-panels \textbf{ii-iv} show the response under two pulses with different spacings: 10 ns (\textbf{ii}), 30 ns (\textbf{iii}), 50 ns (\textbf{iv}).  We show the responses at a particular frequency (3.657 GHz for the AlOx sample (\textbf{a}), and 4.254 GHz for the silicon oxide sample (\textbf{b})), selected due to their long ring-downs. We sweep the amplitude of the first pulse from 0 to 30000 arbitrary units, keeping the amplitude of the second pulse fixed at 10000. All drive amplitudes are calibrated (See Ref. \cite{BCTDS} Appendix E), and the phase of both pulses is fixed. The result is coherent control of the non-Markovian collapse and revival responses from the ensemble of TLS defects. This behavior is observed consistently across both types of samples.
}
\label{fig:spacing_amp} 
\end{figure}
  
The BCTDS technique offers a powerful platform for probing the driven dynamics of TLS defect ensembles across a wide range of materials. Using pulses with sharp rise and fall times, defects are excited through sudden “quenches” that initiate nonadiabatic evolutions. These dynamics can be naturally framed within the Floquet formalism, where carefully chosen pulse parameters and amplitudes can produce destructive interference, strongly suppressing transitions between Floquet states \cite{PhysRevA.94.032323}. Similar effects have been demonstrated in pairs of superconducting qubits to create dark states that decouple to a waveguide environment \cite{Zanner2022}. Such interference control presents a promising route to optimal qubit operation, mitigating the deleterious influence of the surrounding bath of defects. While most prior work has focused on single-spin (qubit) dynamics during gate operations, the collective behavior of defect ensembles—highly relevant in realistic devices—remains less understood. The BCTDS approach is ideally suited to address this gap. In particular, sequences of multiple pulses enable direct observation of bath memory effects: within the observed coherence window of hundreds of nanoseconds, the ensemble retains information about the initial excitation, which then interferes with subsequent pulses. This persistence of memory manifests in collapse-and-revival patterns \cite{BCTDS} and can be interpreted in terms of repeated Landau–Zener transitions between dressed states, where the bath’s coherent back-action plays a decisive role in shaping the system’s evolution.

We investigate the potential prospect of ensemble bath control using pulse arrangements and amplitudes, and perform the following experiment. We use a sapphire sample with a 2 nm ALD aluminum oxide deposit (see Appendix \ref{app_sec:AlOx_prep} for details on sample preparation). The broadband nature of the waveguide allows us to explore the response across a wide frequency range of 3-5 GHz (shown Fig.\,\ref{fig:spacing_amp}\textbf{ai}). To explore the effect of various drive parameters, we select a frequency cut at 3.657 GHz, corresponding to a location of prominent ring-down (indicated by the horizontal dashed line). We examine the transient spectrum following the application of two consecutive pulses sent and measured at this fixed frequency. The pulses are separated by 10 ns, as illustrated in Fig.\,\ref{fig:spacing_amp}\textbf{aii}. The first pulse amplitude is swept from 0 to 30000 a.u. (full power of the RFSoC instrument) to coherently control the memory state of the system and its interference with the second pulse.  The second pulse is held constant at 10000 a.u., with its phase and time position also held the same across all sweeps. Therefore, we keep the second pulse as a constant variable, with any change in the response afterwards strictly a result of the memory affecting the system due to the first pulse. As shown in Fig.\,\ref{fig:spacing_amp}\textbf{aii}, the ring-down patterns exhibit clear shifting and interference changes, modified coherently by the amplitude of the first pulse. We introduce an additional control parameter by varying the time separation between the two pulses, shifting the first pulse earlier in time. The resulting spectra are shown in Fig.\,\ref{fig:spacing_amp}\textbf{aii-iv}, corresponding to separations of 10, 30, and 50 ns, respectively. Comparison between these instances shows systematic shifts and distortions that reflect the evolving influence of the first-pulse memory on the second-pulse response.

In order to investigate whether this response is specific to the particular sample featuring AlOx deposits, we repeat this experiment for a different sample, featuring a silicon sample with a native oxide layer at the surface (same sample used in Fig.\,\ref{fig:phase_V}). The broad spectrum of 3-5 GHz is shown Fig.\,\ref{fig:spacing_amp}\textbf{bi}, where we pick an appropriate frequency with a long coherent ring-down, at 4.254 GHz. The identical dual pulse experiment is performed with qualitatively similar results shown in Fig.\,\ref{fig:spacing_amp}\textbf{bii-iv}. By sweeping the amplitude and temporal position of a strong initial pulse, we observe systematic modifications in the ring-down patterns following a constant-parameter second pulse, attributable to the bath’s memory, which remains coherent over hundreds of nanoseconds.

It is worth clarifying that the exact ring-down features are dependent on the actual system parameters, such as detuning and coupling, as well as the density and spatial distributions of the defects. These parameters are different across the two samples, which is reflected in the different features across the two spectra. For example, in the silicon oxide sample, the transient response at the chosen frequency remains coherent for a shorter duration. We adjust the color scale accordingly to avoid distracting the reader with these differences and to emphasize the underlying dynamics, which are qualitatively similar across samples. This comparison demonstrates that coherent control and targeted manipulation of the bath’s memory using amplitude and inter-pulse spacing can be implemented across different materials and defect environments. 

\begin{figure}[hpbt!]
\begin{centering}
\includegraphics[width=0.5\textwidth]{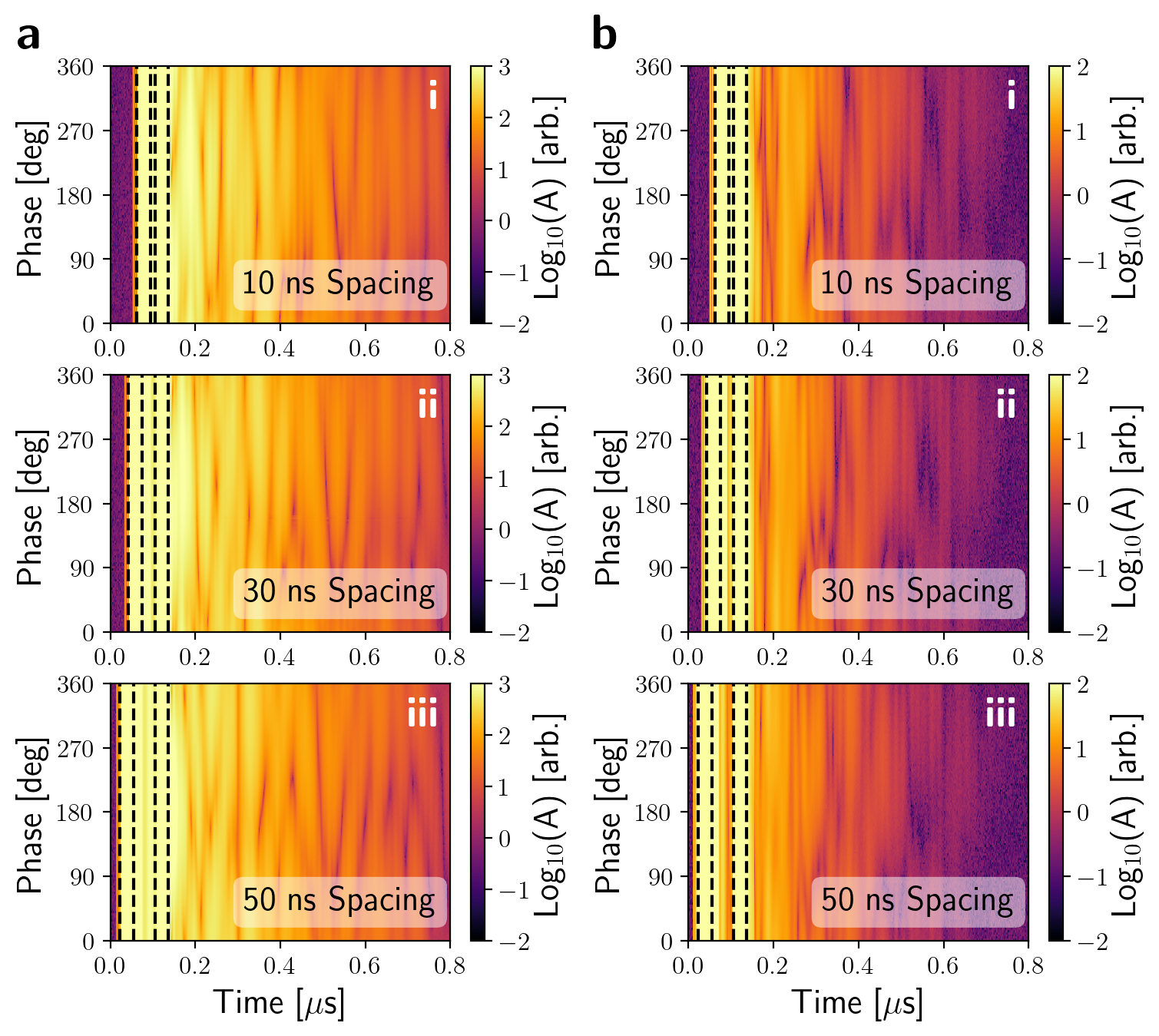}
\end{centering}
\caption{\label{fig:spacing_phase} 
Coherent control of TLS defect ensembles in different samples using pulse phase and spacing, setup is the same as Fig.\,\ref{fig:spacing_amp}. \textbf{a,} Sapphire sample with a thin 2 nm AlOx deposit via ALD. \textbf{b,} Silicon samples with a native oxide layer. Sub-panels \textbf{i-iii} show the response at a frequency cut (3.657 GHz for the AlOx sample (\textbf{a}), and 4.254 GHz for the silicon oxide sample (\textbf{b})) under two pulses with different spacing: 10 ns (\textbf{i}), 30 ns (\textbf{ii}), and 50 ns (\textbf{iii}). The magnitudes of the two pulses are kept unchanged for both samples, with the first pulse being 3 times stronger than the second, and all drive amplitudes are calibrated (see Ref. \cite{BCTDS} Appendix E). We sweep the phase of the first pulse from 0 to 360 degrees, keeping the phase of the second pulse fixed, resulting in coherent control of the non-Markovian responses from the ensemble of TLS defects, similar to Fig.\,\ref{fig:spacing_amp}. This behavior is observed consistently across both types of samples.}
\end{figure}

In addition to pulse amplitude, phase can also serve as a tuning parameter for controlling transient interference patterns. This is illustrated in Fig.\,\ref{fig:spacing_phase} for the same AlOx and silicon samples used in Fig.\,\ref{fig:spacing_amp}. As in the previous amplitude-sweep experiment, the second pulse is held constant, with fixed amplitude (10000 a.u.), phase, and position in time. The initial pulse is sent at the full amplitude of 30000 a.u., and its phase is swept from 0 to 360 degrees, while the inter-pulse spacing is varied, as shown in Fig.\,\ref{fig:spacing_phase}\textbf{i-iii}, for the AlOx sample (\textbf{a}) and silicon sample (\textbf{b}). The resulting spectra exhibit clear shifts and distortions, reflecting coherent manipulation of the bath. Notably, these features are qualitatively consistent across both samples.

\begin{figure}[hpbt!]
\begin{centering}
\includegraphics[width=0.5\textwidth]{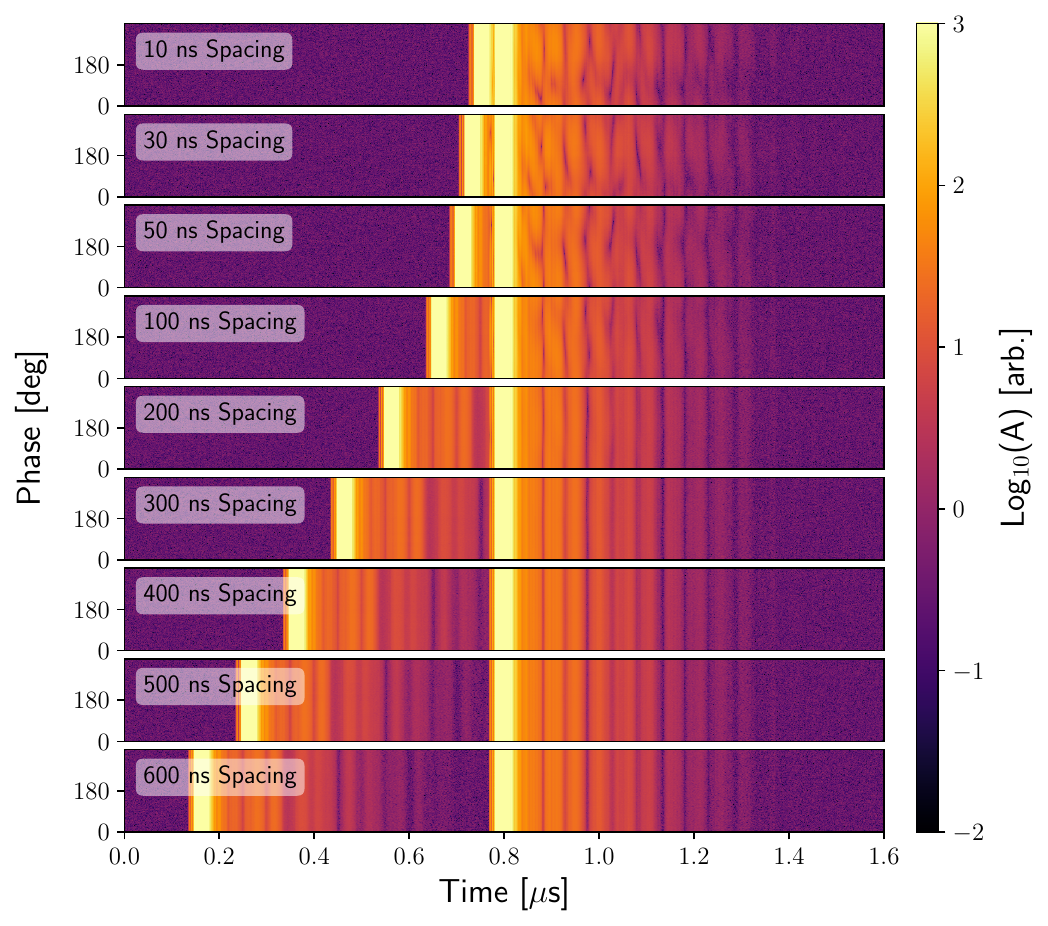}
\end{centering}
\caption{Systematic study of the memory effect as a function of inter-pulse delay, while varying the phase of the first pulse relative to the second. We perform phase coherent control on sapphire samples with spin-coated Shipley 1813 photoresist at 3.4 GHz (similar to Fig.\,\ref{fig:spacing_phase}), sweeping the inter-pulse delay from 10 to 600 ns. For short delays, residual ring-down from the first pulse is still resolvable, indicating incomplete relaxation and a remaining memory that interferes with the second pulse. As the delay increases significantly beyond the extracted decay lifetime of the single pulse ring-down ($\sim$100 ns), the interference patterns diminish and coherent control is lost.}
\label{fig:long_pulse_spacing}
\end{figure}

We further investigate the time scale of this memory effect and  demonstrate the robustness of the experiment across various defect-rich materials in the following analysis. We use sapphire samples with spin-coated Shipley 1813 photoresist, mounted in the same configuration as the silicon and aluminum oxide samples presented above. We perform the phase control experiment at 3.4 GHz, where the inter-pulse delay is varied from a wider range of 10 to 600\,ns (Fig.\,\ref{fig:long_pulse_spacing}). ring-down patterns from the first pulse persist for a finite interval, reflecting memory from previous excitations of the TLS defects. A second pulse within this interval produces strong interference, and inter-pulse spacing serves as an effective control knob that coherently manipulates the transient response. We extract the lifetime of the single pulse response, using a simple exponential fit (\cite{BCTDS}), to be $\sim$100 ns. As the delay increases significantly beyond this lifetime, the interference patterns between the two excitations gradually diminish, showing that the degree of coherent control depends sensitively on the pulse spacing.

These demonstrations show that constructive and destructive interference can be utilized to modify the collective driven dynamics of defect ensembles through pulse-level control knobs such as amplitude, inter-pulse spacing, and phase. The result is applicable to samples that are different in atomistic structures and provides a clear route for coherent control and targeted manipulation of the bath’s memory, informing strategies for mitigating decoherence and harnessing bath dynamics for optimal control. Understanding these response mechanisms not only offers deep insight into the nonlinear driven response of ensemble spin defects and non-Markovian effects, but also leads to improvements to the performance of superconducting circuits and other defect-sensitive platforms.

\section{\label{sec:theory_results} Numerical and Theoretical Results}

\begin{figure}[hpbt!]
\begin{centering}
\includegraphics[width=0.48\textwidth]{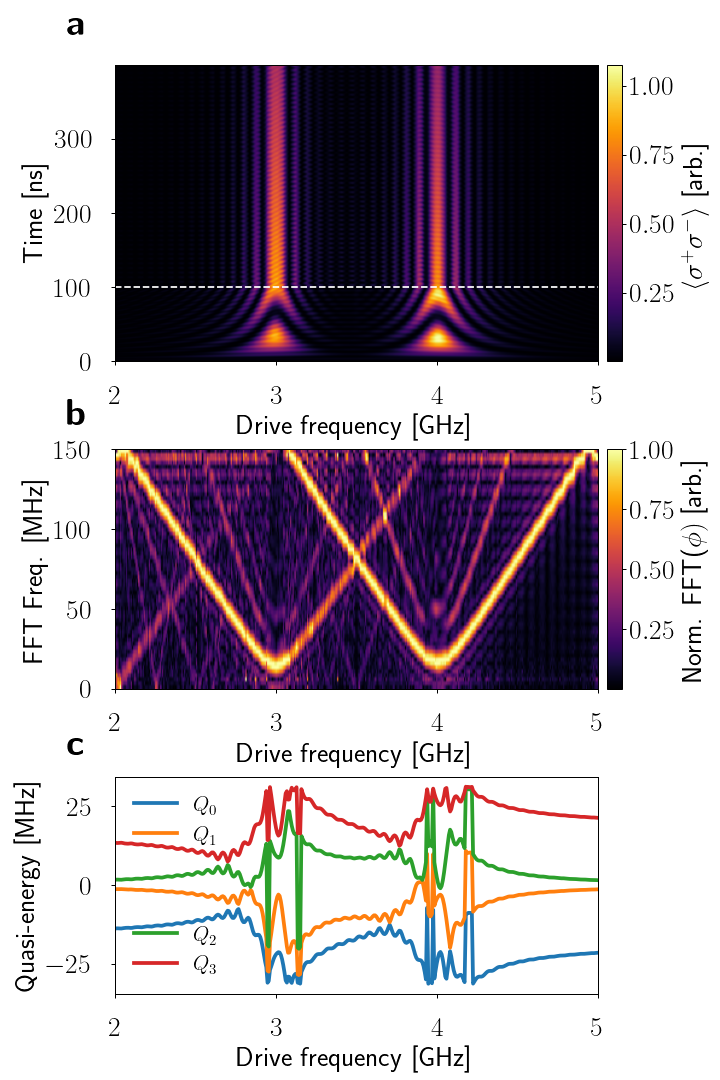}
\end{centering}
\caption{Numerical simulations for the transient response, spectral features, and Floquet quasi-energies for a $N = 2$ spin system. \textbf{a,}  Time-domain evolution of the collective population $\langle \sigma^{+}\sigma^{-}\rangle$, obtained by solving the Lindblad master equation. The simulation employs a square–cosine pulse with dipole–dipole coupling $J/2\pi = 50$ $\text{MHz}$, drive amplitude $A/2\pi = 100$ $\text{MHz}$, pulse duration $\tau = 100$ $\text{ns}$ (horizontal dashed line), collective dissipation rate $\Gamma/2\pi = 2.0$ $\text{MHz}$, and bare transition frequencies $\omega_{1}/2\pi = 3.0$ $\text{GHz}$ and $\omega_{2}/2\pi = 4.0$ $\text{GHz}$. \textbf{b}, Normalized amplitude of the fast-Fourier transform (FFT) of the phase $\phi(t) = \arg(\langle\sigma_1^{+}\rangle / \langle\sigma_2^{+}\rangle)$, calculated from the time-domain data in panel~\textbf{a}. Each spectrum is divided by its maximum value to highlight the dominant frequency components. \textbf{c,} Floquet quasi-energy spectrum extracted from the single-period propagator. The four lowest branches exhibit pronounced features that emerge at drive frequencies where long-lived ring-down oscillations are observed in panel \textbf{a}, as well as the center of the more prominent spectral features observed in \textbf{b}.}
\label{fig:floquet_quasienergies}
\end{figure}

\subsection{\label{subsec:theory_background} Theoretical Background}

In previous sections, we presented experimental results on various samples via BCTDS. Here, we introduce an effective spin model that qualitatively captures the main features observed in the experiment and serves as an interpretative framework. We modeled the system using a standard driven tunneling model with dipole-dipole interactions \cite{deLosSantosSanchez2022}.
The system under study consists of an ensemble of two-level defects embedded in a broadband photonic environment, forming a many-body system of interacting defects driven by an external time-dependent electric field. This external field couples with the TLS dipoles through the interaction term \(\mathbf{E}(t) \cdot \mathbf{d}\), where \(\mathbf{E}\) is the applied electric field and \(\mathbf{d}\) is the dipole moment. Furthermore, the TLS defects interact among themselves via dipole--dipole interactions. To model this, we employ a standard tunneling framework extended to include both external driving and dipole--dipole couplings, enabling computation of the TLS response.

To understand this system, we first consider a single TLS defect characterized by an energy detuning \(\varepsilon_j\) and a tunneling amplitude \(\Delta_j\), as illustrated in Fig.\,\ref{fig:overview}\textbf{a}. The effective Hamiltonian for the \(j\)-th TLS defect in the position basis \(\{|L\rangle, |R\rangle\}\) can be written as \cite{Lisenfeld2015}:
\begin{equation}
\hat{H}_j = \frac{1}{2} \begin{pmatrix} \varepsilon_j & \Delta_j \\ \Delta_j & -\varepsilon_j \end{pmatrix} 
= \frac{1}{2} \left(\varepsilon_j \hat{\sigma}_z^{(p,j)} + \Delta_j \hat{\sigma}_x^{(p,j)}\right),
\label{eq:tls_position_basis}
\end{equation}
where \(\hat{\sigma}_z^{(p,j)}\) and \(\hat{\sigma}_x^{(p,j)}\) are Pauli matrices in the position basis representing the difference between left and right localized states. Diagonalizing this Hamiltonian yields the energy eigenstates with eigenenergies 
\begin{equation}
E_j = \sqrt{\varepsilon_j^2 + \Delta_j^2},
\label{eq:tls_energy_splitting}
\end{equation}
with a mixing angle \(\theta_j\) defined through $\tan \theta_j = \Delta_j/\varepsilon_j$. Transforming to this energy eigenbasis, the static Hamiltonian simplifies to
\begin{equation}
\hat{H}_{0,j} = \frac{1}{2} E_j \hat{\sigma}_z^{(j)},
\label{eq:tls_energy_basis}
\end{equation}
where \(\hat{\sigma}_z^{(j)}\) now acts in the energy eigenbasis.

Next, we introduce the interaction of the ensembles of TLS defects with an external, time-dependent electric field \(\mathbf{E}(t)\). The coupling occurs through the electric dipole moment \(\mathbf{p}_j\) of the TLS defects. Choosing the direction of the dipole moment as the quantization axis (here aligned along \(\hat{\mathbf{x}}\)), the collective polarization operator for an ensemble of \(N\) TLS defects can be defined as
\begin{equation}
\hat{\mathbf{P}} = \sum_{j=1}^N p_j \left( \cos \theta_j \hat{\sigma}_z^{(j)} + \sin \theta_j \hat{\sigma}_x^{(j)} \right) \hat{\mathbf{x}},
\label{eq:collective_polarization}
\end{equation}
where the Pauli operators are in the energy eigenbasis. The full Hamiltonian, including the interaction with the electric field, and dipole-dipole interactions, is then written as
\begin{equation}
\hat{H}(t) = \sum_{j=1}^{N} \hat{H}_{0,j} - \hat{\mathbf{P}} \cdot \mathbf{E}(t) + \sum_{i < j} J_{ij} \hat{\sigma}_x^{(i)} \hat{\sigma}_x^{(j)},
\label{eq:hamiltonian}
\end{equation} where $\mathbf{E}(t) = A(t)\cos(\omega_d t) \hat{\mathbf{x}}$, with $A(t)$ equal to $A_0$ for $0 \le t \le \tau$ and zero otherwise. In our experiments, $\tau$ represents the duration of the applied drive. Moreover, dissipation into a broadband waveguide is modeled by collective decay with jump operator $\hat{S}_\pm = \sum_{j=1}^N \hat{\sigma}_\pm^{(j)}$. The dynamics obey the Lindblad master equation
\begin{equation}
\frac{d\rho}{dt} = -i \left[ \hat{H}(t), \rho \right] + \Gamma \left( 2 \hat{S}_- \rho \hat{S}_+ - \{\hat{S}_+ \hat{S}_-,\rho \} \right),
\end{equation}
where $\hat{S}_+ = \hat{S}_-^{\dagger}$ and $\Gamma$ is the collective decay rate. The assumption of collective decay is motivated by the broadband nature of the waveguide, which couples similarly to all TLS defects.

To this end, we model the dynamics of interacting driven-spins subject to a finite-duration periodic pulse. This provides a simplified framework for interpreting the interference patterns observed in the TLS ring-down and understanding how the applied drive shapes the dynamics. Since the drive is periodic in this region, we used Floquet to compute quasi-energies within the drive duration. Given that the system absorbs energy from the drive in this region and re-emits it when the driving is turned off, we expect to see transient ring-downs accompanied by sidebands. After the end of the pulse, signatures of the prior Floquet dynamics may persist in the post-pulse evolution, manifesting as coherent oscillations or ring-down features. To support this interpretation, we briefly summarize the Floquet formalism in Appendix \ref{app_sec:floquet}, which we employed to numerically simulate and interpret the behavior of the system in the driven regime.

\subsection{Floquet Quasi-Energies}
\label{sec:floquet}
\begin{figure}[htbp!]
\begin{centering}
\includegraphics[width=0.5\textwidth]{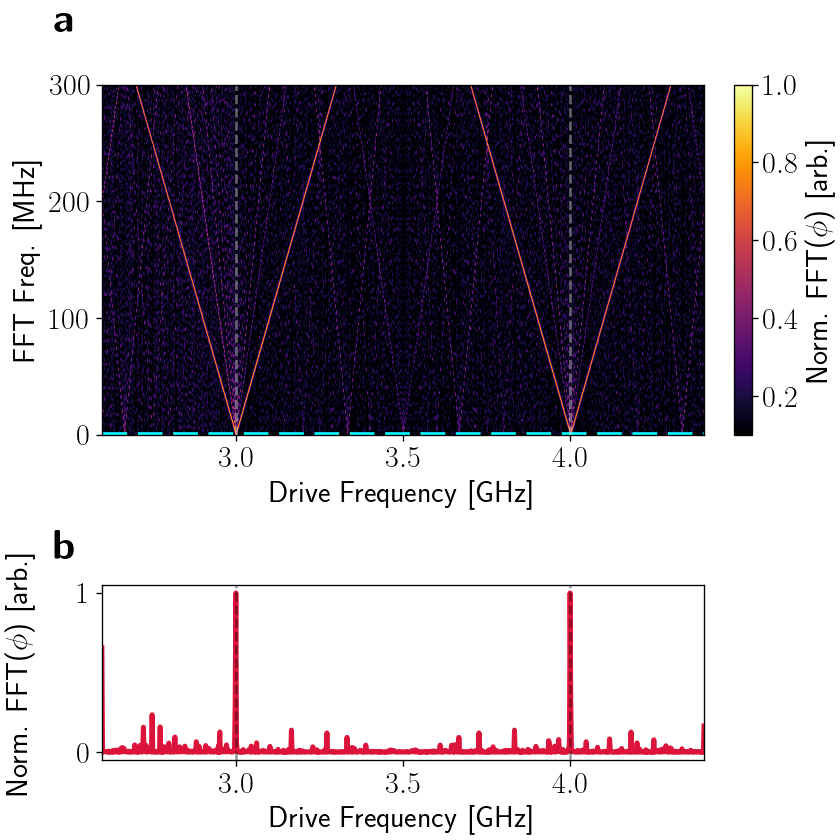}
\end{centering}
\caption{Analytical spectral signatures of the collective evolution.
\textbf{a,} Normalized fast-Fourier transform of the relative phase for two coupled two-level systems ($\omega_{1}/2\pi=3.0\,\text{GHz}$, $\omega_{2}/2\pi=4.0\,\text{GHz}$).
\textbf{b,} Zero-frequency (DC) component of FFT ($\phi$) of the analytical spectrum. Sharp peaks at the bases of the dominant V-shaped branches mark the primary resonances, while weaker lines arise from sideband dressing of the eigenmodes.}
\label{fig:analytical_vs}
\end{figure}

Fig.\,\ref{fig:floquet_quasienergies} shows numerical simulations that qualitatively reproduce the experimentally observed phase V pattern in the $I$–$Q$ plane (Fig.~\ref{fig:phase_V}\textbf{a}).  
Fig.\,\ref{fig:floquet_quasienergies}\textbf{a} presents results from a full master equation simulation including the complete pulse sequence and dissipation. Fig.\,\ref{fig:floquet_quasienergies}\textbf{b} shows the FFT of the relative phase accumulated by the TLS defects, while Fig.~\ref{fig:floquet_quasienergies}\textbf{c} displays the Floquet quasi-energy spectrum calculated during the periodic drive, highlighting the regions of sharp spectral features.  Notably, the characteristic V-shaped features observed in the experiment (Fig.~\ref{fig:phase_V}\textbf{c}) are clearly visible in the simulated spectra of panel~\textbf{b} and coincide with dips in the quasi-energy branches in panel~\textbf{c}. These results indicate that the phase V structure originates from interference between Floquet-driven modes, which hybridize the system eigenstates during the drive.

\subsection{Analytical Expression from Second Order Perturbation Theory}

To understand these features analytically, we consider two coupled TLS defects restricted to the single-excitation subspace spanned by $\{\ket{eg}, \ket{ge}\}$. The TLS defects are coupled with strength \(J\) and driven by a time-dependent field of amplitude \(\Omega\). In the weak-driving limit, we treat the drive as a perturbation and use time-dependent perturbation theory to compute the excitation amplitude of the first TLS defect

\begin{equation}
\begin{aligned}
c_{eg}(t) &= \frac{\Omega}{2} e^{-i \omega_1 t} \frac{1 - e^{i \delta_1 t}}{\delta_1} \\
&\quad + \frac{\Omega}{2} J e^{-i \omega_1 t} \frac{1}{\delta_2} \left( \frac{1 - e^{i \delta_1 t}}{\delta_1} + \frac{1 - e^{-i \delta t}}{\delta} \right) \\
&\quad + i \frac{\Omega}{2} J^2 e^{-i \omega_1 t} K,
\end{aligned}
\end{equation}
where \(\delta = \omega_1 - \omega_2\), $\delta_1 = \omega_1 - \omega_d$, $\delta_2 = \omega_2 - \omega_d$ and

\begin{equation}
\begin{aligned}
K &= \frac{1}{i \delta_1} \biggl\{ \frac{1}{i (\delta + \delta_1)} \left[ \frac{e^{i \delta_1} - 1}{i \delta_1} - \frac{e^{-i \delta t} - 1}{-i \delta} \right] \\
&\quad - \frac{1}{i \delta} \left[ t - \frac{e^{-i \delta t} - 1}{-i \delta} \right] \biggr\}.
\end{aligned}
\end{equation}

Physically, each term captures different dynamical contributions.
The first term describes direct off-resonant driving of the first TLS at frequency $\omega_{d}$, with detuning-dependent phase accumulation, analogous to Ramsey interference. The second term, proportional to $J$, captures virtual excitation exchange between the two TLS defects and includes interference factors such as $1/(\delta_1 \delta_2)$ and $1/(\delta_2\delta)$, which arise from multiple indirect excitation paths. 
The third term, proportional to $J^{2}$, accounts for higher-order virtual exchange processes; the integral $K$, summing phase contributions from sequences of inter-TLS hopping. Fig.\,\ref{fig:analytical_vs}\textbf{a} shows the spectral response predicted by this analytical expression for the coherence dynamics.  Fig.\,\ref{fig:analytical_vs}\textbf{b} shows the zero-frequency (DC) component along the cyan trace in \ref{fig:analytical_vs}\textbf{a}. Two prominent peaks appear at the bases of the V-shaped features centered at the bare resonances $\omega_{1}$ and $\omega_{2}$. Additional weaker peaks reflect sideband dressing and higher-order collective processes. These trends are consistent with the numerical results in Fig.~\ref{fig:floquet_quasienergies}\textbf{b} and with the experimental measurements in Fig.~\ref{fig:phase_V}\textbf{c–d}.

\subsection{Amplitude Control}
\begin{figure}[hpbt!]
\begin{centering}
\includegraphics[width=0.5\textwidth]{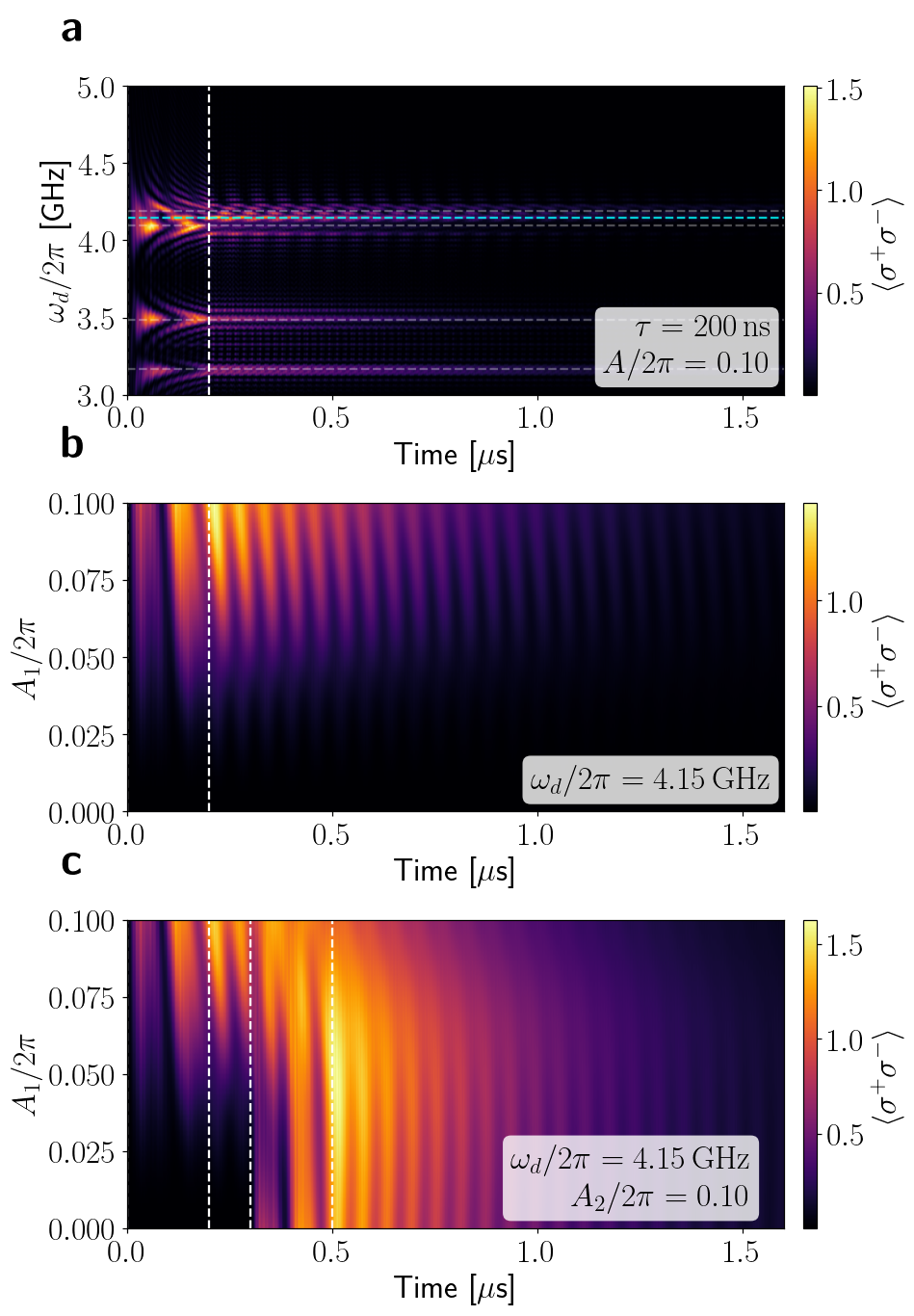}
\end{centering}
\caption{Amplitude and multi-pulse control of a TLS ensemble.
\textbf{a,} Collective excitation for a system size of $N=4$ TLS defects following a single square pulse of duration $\tau = 200$ ns and amplitude $A_{1}/2\pi=100$ MHz. Dipole–dipole couplings are sampled uniformly from $J/2\pi \in [-50,50]$ MHz, and bare resonance frequencies from $\omega_i/2\pi \in [3.0,5.0]$ GHz. The resulting resonance frequencies are $\omega_1/2\pi=3.49$ GHz, $\omega_2/2\pi=3.17$ GHz, $\omega_3/2\pi=4.09$ GHz, and $\omega_4/2\pi=4.19$ GHz, while the collective decay rate is $\Gamma/2\pi= 2.0$ MHz. The cyan marker denotes the drive frequency that maximizes the ring-down, $\omega_{d}/2\pi=4.15$ GHz, used in \textbf{b,} and \textbf{c}. \textbf{b,} Amplitude sweep at the fixed frequency $\omega_{d}/2\pi$ reveals collapse-and-revival behavior in the ring-down that intensifies with increasing pulse amplitude. \textbf{c,} Two-pulse protocol, namely a first pulse of variable amplitude $A_{1}$ is followed, after a gap $\tau_{g}=100$\,ns, by a second pulse of fixed amplitude $A_{2}/2\pi=100$\,MHz. Each pulse lasts $\tau = 200$ ns. Interference and memory effects arise from the overlap between residual oscillations of the first pulse and the excitation induced by the second at subsequent times.
}\label{fig:theo_amplitude_control}
\end{figure}

Fig.\,\ref{fig:theo_amplitude_control} demonstrates how coherent control can be achieved in a system of four weakly interacting TLS defects using pulsed driving, revealing rich interference dynamics and long-lived memory effects. In Fig.\,\ref{fig:theo_amplitude_control}\textbf{a}, a short square-envelope drive pulse of fixed amplitude and duration is applied while sweeping the drive frequency. After the drive is switched off (marked by the white dashed line), the system exhibits pronounced ring-down oscillations at distinct frequencies corresponding to the bare transition energies of the individual TLS defects, indicating that energy absorbed during the pulse persists as coherent population oscillations at subsequent times. The fact that these ring-downs are sharp and long-lived reveals that the system retains phase coherence, with the interaction between TLS defects enabling a collective response. The structure and amplitude of the ring-down depend on how the drive frequency overlaps with the coupled modes, when the drive is near-resonant with one or both TLS defects, constructive interference enhances excitation. Off-resonant driving, however, results in weaker or destructively interfering excitation pathways.

In Fig.~\ref{fig:theo_amplitude_control}\textbf{b}, the drive frequency is fixed near a resonance identified in panel~\textbf{a} by the cyan line at $\omega_{d}/2\pi = 4.15$\,GHz, and the amplitude $A_{1}$ is varied. Increasing $A_{1}$ enhances the dressing of the TLS defects during the pulse, resulting in stronger excitation and more pronounced post-pulse ring-downs. This amplitude dependence reflects how hybridization during the drive modifies the initial state of the system prior to free evolution. After the drive is switched off, the dressed states interfere, producing the observed ring-down profile in which collapse-and-revival oscillations become increasingly pronounced at higher amplitudes. Fig.\,\ref{fig:theo_amplitude_control}\textbf{c} also considers a second amplitude component, $A_{2}$, kept fixed while varying $A_{1}$. This configuration reveals memory effects in the TLS bath, arising from interference between excitations generated by the first pulse and those induced by the second pulse after a delay $\tau_{g} = 100$ ns. The two-pulse drive enables multi-path interference between Floquet sidebands, producing conditions where excitation is either enhanced or suppressed depending on the accumulated relative phase between the drive components and TLS modes. The resulting vertical bands in the time–amplitude maps correspond to regions of constructive and destructive interference, providing a means to selectively control the collective relaxation dynamics.

Overall, these results demonstrate that the relaxation dynamics of the system are governed not only by the drive but also by interference between dressed eigenstates and the memory preserved in their coherent evolution. The behavior in Fig.~\ref{fig:theo_amplitude_control} is qualitatively consistent with the experimental observations in Fig.~\ref{fig:spacing_amp}.

\subsection{Separation and Phase Control}
\begin{figure}[hpbt!]
\begin{centering}
\includegraphics[width=0.5\textwidth]{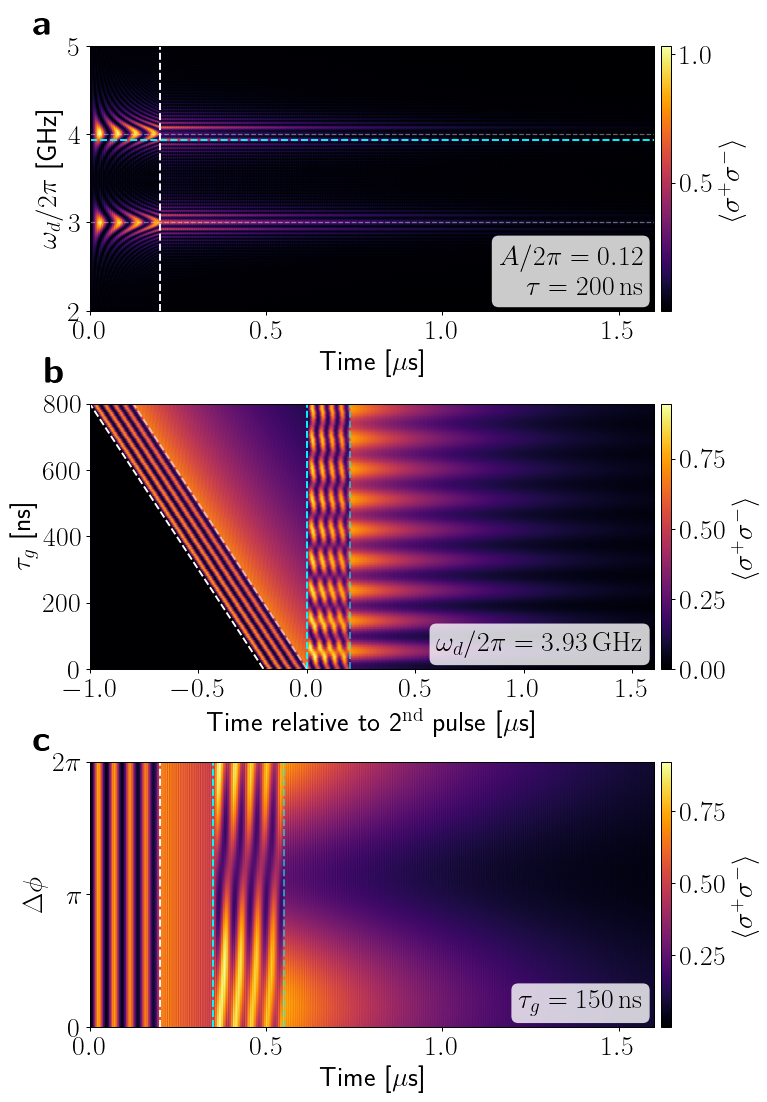}
\end{centering}
\caption{Coherent control of a two-TLS ensemble. \textbf{a,} Time-domain evolution for $N = 2$ coupled TLS defects, with interaction strength $J/2\pi = 50$ MHz. The cyan marker indicates the drive frequency that maximizes the ring-down, $\omega_{d}/2\pi=3.93$ GHz, used in panels \textbf{b} and \textbf{c}. The square pulse has a duration of $\tau = 200$ ns, amplitude of $A/2\pi = 120$ MHz, and collective dissipation rate $\Gamma/2\pi = 2.0$ MHz.  
\textbf{b,} Dependence of the collective response on the inter-pulse gap $\tau_{g}$. The color map shows the excitation probability versus $\tau_{g}$, and time measured from the start of the second pulse. Interference fringes fade as $\tau_{g}$ increases, indicating reduced overlap between the two ring-down signals mainly due to collective dissipation of the TLS ensemble. \textbf{c,} Relative-phase control with a fixed gap $\tau_{g}=150$ ns. Sweeping the phase difference $\Delta\phi$ between two identical pulses ($\tau = 200$ ns, $A/2\pi = 120$ MHz) produces clear constructive and destructive interference, modulating $\langle\sigma^{+}\sigma^{-}\rangle$ from near-complete revival to strong suppression at intermediate $\Delta \phi$ values.} \label{fig:separation_phase_control}
\end{figure}

To further investigate coherent control in a weakly interacting two-level system of size $N=2$, we simulate a square-ramped drive and monitor the subsequent ring-down dynamics, as shown in Fig.~\ref{fig:separation_phase_control}. Panel~\textbf{a} shows long-lived oscillations after the drive, with frequencies matching the bare transition energies of the TLS defects. This indicates coherent population transfer in the presence of collective single-photon loss and minimal dephasing. 

\begin{figure*}[htpb!]
  \centering 
\includegraphics[width=\textwidth,keepaspectratio]{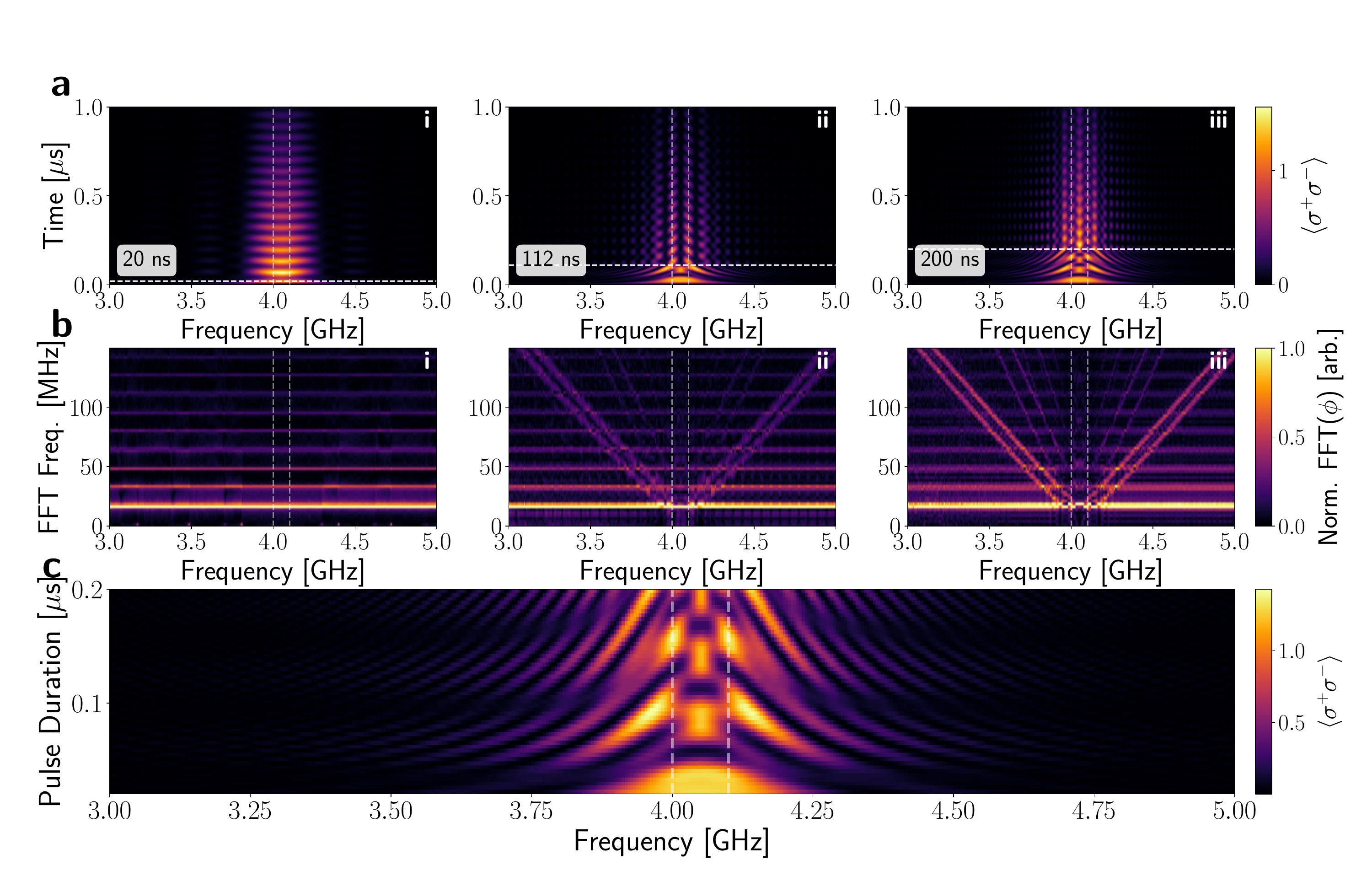}
\caption{Sharpening of ring-downs with increasing pulse duration.
\textbf{a,} Simulated population dynamics $\langle\sigma^{+}\sigma^{-}\rangle$ for pulse durations $\tau$ of 20 ns \textbf{(i)}, 112 ns \textbf{(ii)}, and 300 ns \textbf{(iii)} . The system consists of $N = 2$ TLS with individual resonance frequencies $\omega_1/2\pi = 4.0$ GHz and $\omega_2/2\pi = 4.1$ GHz (horizontal white dashed lines), coupled via a dipole-dipole interaction of strength $J/2\pi = 50$ MHz. The drive amplitude is $A/2\pi = 100$ MHz, and the collective dissipation rate is $\Gamma/2\pi = 2.0$ MHz. \textbf{b}, Spectra of the phase evolution for a two-spin ensemble, $\phi(t) = \arg\left(\langle \sigma_1^{+} \rangle / \langle \sigma_2^{+} \rangle \right)$, showing characteristic V-shaped features centered at the resonance frequencies of the individual TLS defects. These features become more pronounced, and additional spectral components emerge as the pulse duration increases. \textbf{c}, Corresponding population values sampled immediately after the pulse is turned off for all simulated pulse durations. Increasing $\tau$ leads to sharper transient and spectral features, along with the emergence of additional dressed sidebands.}
\label{fig:sharpening_physics}
\end{figure*}

Fixing the drive frequency to the value that produces the most pronounced ring-down, $\omega_d/2\pi = 3.93$ GHz, and we vary the delay, $\tau_g$, between two pulses while keeping their duration and amplitude fixed, as shown in Fig.~\ref{fig:separation_phase_control}\textbf{b}. The resulting oscillatory modulation and interference effects are most visible during the second pulse, which becomes less prominent as the inter-pulse delay systematically increases. This behavior reveals collective memory effects in the TLS bath, consistent with Floquet quasi-energy splitting and dressing of Floquet modes, which are further manifested in the enhanced ring-down oscillations following the second pulse. Panel~\textbf{c} explores this regime by introducing a second tone at fixed separation and sweeping the relative phase, $\Delta \phi$, between the tones. The observed changes in the coherent response demonstrate precise phase control of the collective excitation and population dynamics, further highlighting memory effects and interference control arising from the coherent nature of the system.

These results show that even in a weakly interacting regime, two coupled TLS defects can display nontrivial Floquet dynamics and memory-enhanced control through multi-tone interference. The trends in Fig.~\ref{fig:separation_phase_control} are qualitatively consistent with the experimental results in Fig.~\ref{fig:spacing_phase}.

\subsection{Effect of Pulse Duration on Spectral Features}

Fig.\,\ref{fig:sharpening_physics}\textbf{a} shows the simulated population dynamics for fixed pulse durations $\tau$ = 20 ns,  112 ns, and  200 ns. Increasing the pulse duration results in narrower spectral bandwidths, which in turn sharpen the ring-downs and reveal finer spectral features \cite{BCTDS}. The corresponding phase spectra, obtained from $\phi(t) = \arg(\langle\sigma_1^{+}\rangle / \langle\sigma_2^{+}\rangle)$ over the full time evolution, are shown in Fig.\,\ref{fig:sharpening_physics}\textbf{b}. These maps exhibit characteristic V-shaped features centered at the resonance frequencies of the individual TLS defects, which become more pronounced as $\tau$ increases. Longer pulses also reveal higher-order spectral components, reflecting the dressing of additional sidebands. The corresponding population values sampled immediately after the pulse ends, for all simulated pulse durations, are shown in Fig.\,\ref{fig:sharpening_physics}\textbf{c}. Here, the sharpening of the main resonance and the progressive appearance of sidebands with increasing $\tau$ are clearly visible. To highlight these effects, the two TLS defects were chosen to be close in frequency, enhancing interference between sidebands while sharpening the main resonance as $\tau$ increases. \emph{The simulations also display fast oscillations at short timescales that are absent in the experiments, due to the low-pass filtering inherent to the on-board homodyne detection}. Nevertheless, the overall dynamical signatures with increasing pulse duration are in qualitative agreement with the measurements in Fig.\,\ref{fig:phase_V}\textbf{a, c, d}.

\section{\label{sec:discussion}Discussion}

In this work, we have demonstrated the coherent control of a TLS defect ensemble over a broadband spectrum. Our results show that it is possible to manipulate both the time-domain evolution of the spin defects and its associated spectral features, which display distinctive V-shaped spectral lines. The time evolution of the relative phase between the quadrature signals reveals sharper, better-resolved spectral features than the FFT of the amplitude spectrum. Here, we implement multiple control parameters, including pulse duration, pulse amplitude, and inter-pulse delay, as well as the relative amplitude and phase between two pulses. Through multi-pulse control, we also confirm the presence and tuning of memory effects in the TLS defect ensemble.

These findings have direct relevance for quantum information science. The prolonged ring-down times observed here occur on timescales comparable to inter-pulse delays in gate sequences used in state-of-the-art error correction protocols. Over these timescales, TLS defect relaxation dynamics can interact with the evolution of superconducting qubits, hindering quantum information processing. The memory effects we observe suggest opportunities for optimal control approaches aimed at suppressing TLS-induced decoherence. In principle, machine-learning-based methods could be developed to design pulse sequences that, through careful tuning of amplitude and phase, result in destructive interference in the TLS defect dynamics. Such strategies could substantially attenuate broadband TLS defect ring-downs near the qubit operating frequency, thereby improving qubit coherence and control.

By incorporating the time-dependent drive term into the standard tunneling model, we uncover rich dynamics arising from the dressing of the TLS defect ensembles. These manifest as sidebands that interfere during the transient relaxation, providing insight into the role of TLS–TLS defect interactions. Such interference patterns may also offer a platform for probing excitations in topological materials \cite{Giovannini_2020}. In the low-power regime, we observe weakly dressed modes that reveal the spectral signatures of individual, weakly coupled TLS defects. These fingerprints provide material-specific information and could be used to guide TLS defect mitigation strategies.

BCTDS can be applied throughout fabrication stages without fully packaged qubits, avoiding the narrowband limits of qubit-based probes. It reveals spectral and dynamical features that qubit measurements cannot access. It also enables coherent control of the TLS bath, providing direct access to the collective dynamics that govern defect-induced dissipation. This capability makes it possible to carry out longitudinal studies across materials and process steps to identify universal dielectric response signatures relevant to quantum technologies and fundamental science. The approach extends beyond superconducting circuits to technologies where defects limit performance, including optoelectronic devices, and to platforms such as topological insulators by linking exotic quantum states to their relaxation dynamics. As an outlook, integrating our measurements with first-principles DFT calculations of amorphous structures to estimate interaction strengths, together with effective models for macro-spin ensembles, will provide a full multi-scale description. BCTDS therefore offers a robust platform to probe TLS defects, to coherently manipulate their bath dynamics, and to guide fabrication strategies that mitigate their impact while offering broader applicability in superconducting circuits and beyond.

\section*{\label{sec:data-availability} Data Availability Statement}

The data underlying this article and the code used to generate the figures are available at
\href{https://github.com/fitz-lab/BCTDS-Coherent-Control-Manuscript-Codebase}{\url{https://github.com/fitz-lab/BCTDS-Coherent-Control-Manuscript-Codebase}}.

\begin{acknowledgments}
Startup funds from the Thayer School of Engineering, Dartmouth College, supported this work. We gratefully acknowledge support from DARPA Young Faculty Award No.\, D23AP00192. 
M.O. and V.F. acknowledge startup funds from Cornell University. Partial funding for shared facilities used in this prototype was provided by the Microelectronics Commons Program, a DoD initiative, under award number N00164-23-9-G061.

\end{acknowledgments}

\appendix

\section{Sample Preparation}\label{app_sec:sample_prep}
In section \ref{sec:experimental_results}, we report measurements on two samples: silicon (111) with a native oxide layer and sapphire with aluminum oxide deposited via ALD. Here, we cover the details of the processing steps for both samples.

\subsection{Silicon (111) Sample with a Native Oxide Layer}\label{app_sec:waferpro111_prep}
We use a high-resistivity float zone silicon (111) wafer purchased from WaferPro and first perform a two-step RCA clean: 10 minutes in 6:1:1 {H\textsubscript{2}O:NH\textsubscript{4}OH:H\textsubscript{2}O\textsubscript{2}} at 70~$^\circ$C and 10 min in 6:1:1 {H\textsubscript{2}O:HCl:H\textsubscript{2}O\textsubscript{2}} at 70~$^\circ$C, with water rinses between each step. We then loaded the wafer into a 1000~$^\circ$C furnace for dry oxidation, developing 140 nm of {SiO\textsubscript{x}} on the surface.  We removed the surface oxide layer and prepared H-terminated Si surface with a 5-minute 10:1 BOE bath, water rinse, a 20-minute {NH\textsubscript{4}F} bath, and a water rinse.  Finally, we coated the wafer with a protective resist layer and diced it into 27.5 x 5.5 mm rectangular strips with a dicing saw for shipment. After shipment, the resist layer is removed using AZ 300T Photoresist stripper, followed by an overnight acetone bath at 70~$^\circ$C.

\subsection{Sapphire Sample with Aluminum Oxide Deposit}\label{app_sec:AlOx_prep}
We use a high-grade crystalline sapphire wafer purchased from Crystal Systems. The wafer was coated with AZ1518 photoresist before being diced into 27.5 x 5.5 mm rectangular strips. After dicing, we cleaned the samples using an overnight acetone bath at 70~$^\circ$C and sonication, and then dried them with compressed nitrogen gas. Finally, we deposited a 2 nm AlOx layer onto the cleaned sapphire surfaces using an ALD machine.

\section{Temperature Dependence of BCTDS}
\label{app_sec:temperature_dependence}

\subsection{Temperature Dependence of Ring-down Lifetimes}
\label{app_sec:temperature_dependence_ring_down}
\begin{figure*}[hpbt!]
\includegraphics[width=0.9\textwidth, keepaspectratio]{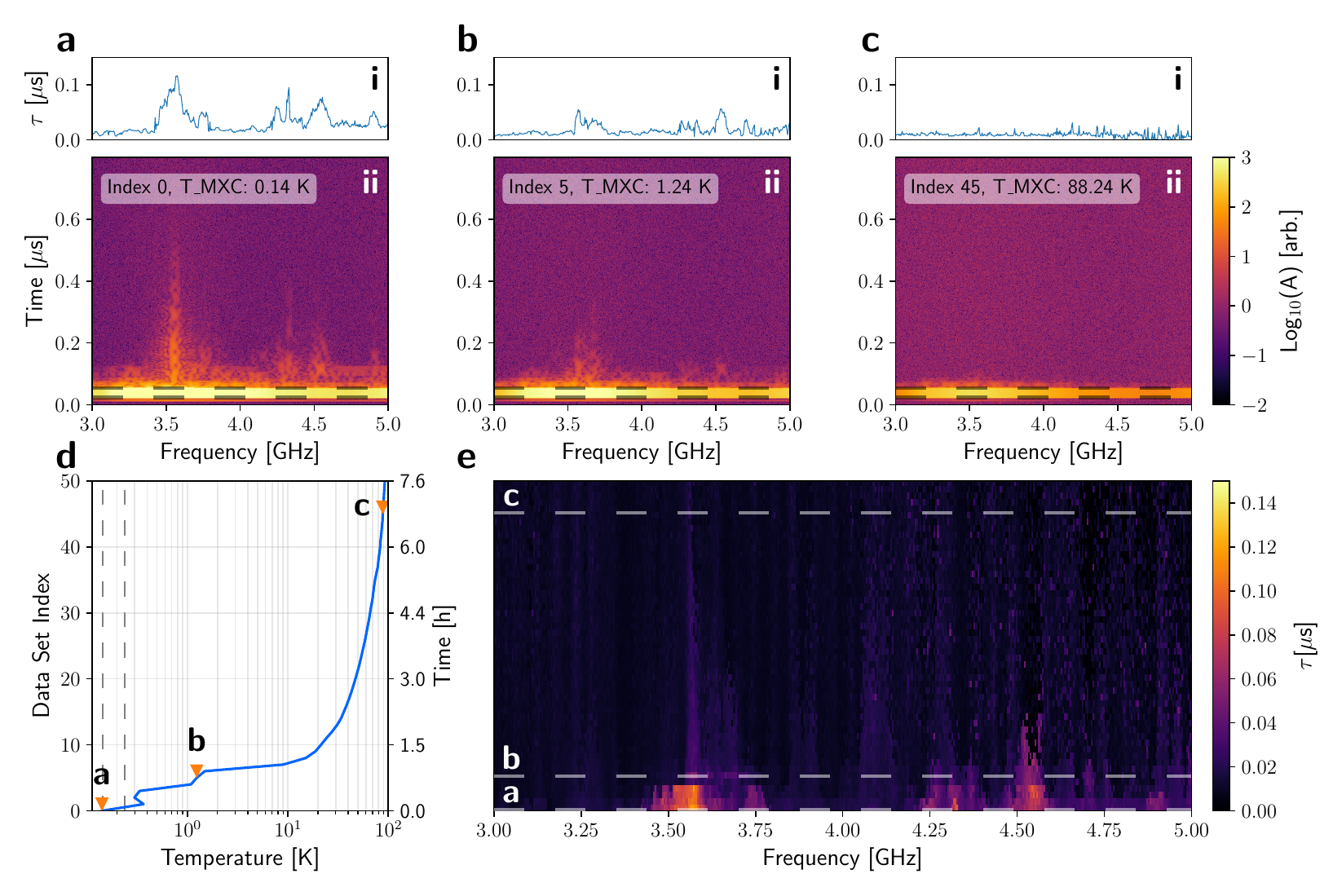}
\caption{Temperature dependence of BCTDS for sapphire samples with spin-coated Shipley 1813 photoresist. We record back-to-back sweeps indexed 0 to 50 during a standard BlueFors warm-up process. \textbf{a-c}, Representative transient spectra (\textbf{ii}) and fitted decay lifetimes $\tau$ (\textbf{i}) at three different temperatures: 0.14 K (index 0), 1.24 K (index 5), and 88.24 K (index 45), where we report the MXC temperature averaged across each sweep. \textbf{e,} Compiled color map of fitted lifetimes $\tau$ for all sweeps (0–50) as a function of frequency, where we mark the representative slices \textbf{a–c} with dashed white lines. \textbf{d,} MXC temperature and time since the warmup started. Slices \textbf{a–c} are marked with orange triangles. We expect the actual sample temperature to be lower than the marked locations due to relatively low thermal conductivity between the baseplate and our insulating sapphire samples. In \textbf{d}, vertical dashed lines at 144 mK and 240 mK indicate the $k_BT=hf$ temperatures for 3 GHz and 5 GHz, which define the lower and upper limits of our sweep range. We observe sharp ring-downs and pronounced collapse-and-revival patterns near base temperature, where quantum coherence effects are most evident. As the temperature increases toward $\sim$1 K, spectral diffusion may lead to slight drifts in the eigenfrequencies, and thermal excitations progressively suppress the coherent response. At temperatures $k_BT \gg hf$, all coherence is lost, and the transient signal disappears.}
\label{fig:temp_dependence}
\end{figure*}

To study the temperature dependence of BCTDS, we perform measurements on sapphire samples with spin-coated Shipley 1813 photoresist during a standard BlueFors warm-up sequence. A heater at the 4K stage rapidly warms up the cryostat, and we record back-to-back BCTDS sweeps, indexed 0 to 50. Figure\,\ref{fig:temp_dependence}\textbf{a-c} shows the representative transient spectra and fitted decay lifetimes at three different temperatures: 0.14 K (index 0), 1.24 K (index 5), and 88.24 K (index 45). We report the mixing chamber (MXC) temperature readings averaged across each sweep, with the MXC sensor located closest to the waveguide and providing the most accurate estimate of the sample temperature. Figure,\ref{fig:temp_dependence}\textbf{e} shows the evolution of the ring-down lifetimes across all dataset indices (0–50) acquired during the warm-up, compiled into a single color map. The corresponding temperature readings and time since the start of the warm-up are shown in Fig.\,\ref{fig:temp_dependence}\textbf{d}. We mark the location of representative slices (Fig.,\ref{fig:temp_dependence}\textbf{a–c}) in the color map, as orange triangles in Fig.,\ref{fig:temp_dependence}\textbf{d} and as horizontal white dashed lines in Fig.,\ref{fig:temp_dependence}\textbf{e}. Due to relatively low thermal conductivity between the base-plate and our insulating sapphire samples, the actual sample temperature is expected to be lower than the average MXC temperature. For reference, we also include two vertical dashed lines in Fig. \ref{fig:temp_dependence}\textbf{d} at 144 mK and 240 mK, corresponding to the thermal energies $k_BT = hf$ for 3 GHz and 5 GHz, respectively. Close to base temperature, the transient response exhibits pronounced ring-downs with long decay times, indicating the presence of coherent TLS defect dynamics. As the temperature approaches 1 K, these features may shift in location and weaken substantially, and the extracted lifetimes decrease sharply. At higher temperatures near 100 K, the transient response is completely suppressed. The absence of response at elevated temperatures indicates that the underlying dynamics are likely of quantum origin, rather than spurious incoherent modes. These observations highlight the strong thermal sensitivity of the TLS ensemble. As temperature increases, the local environment of individual defects changes \cite{Lisenfeld2015}, leading to shifts in their resonance frequencies and coupling strengths that collectively modify the ensemble eigenfrequencies. Ultimately, thermal energy suppresses coherent behavior for high enough temperatures.

\subsection{Temperature Dependence of Coherent Phase Control}

\begin{figure*}[hpbt!]
\includegraphics[width=\textwidth, keepaspectratio]{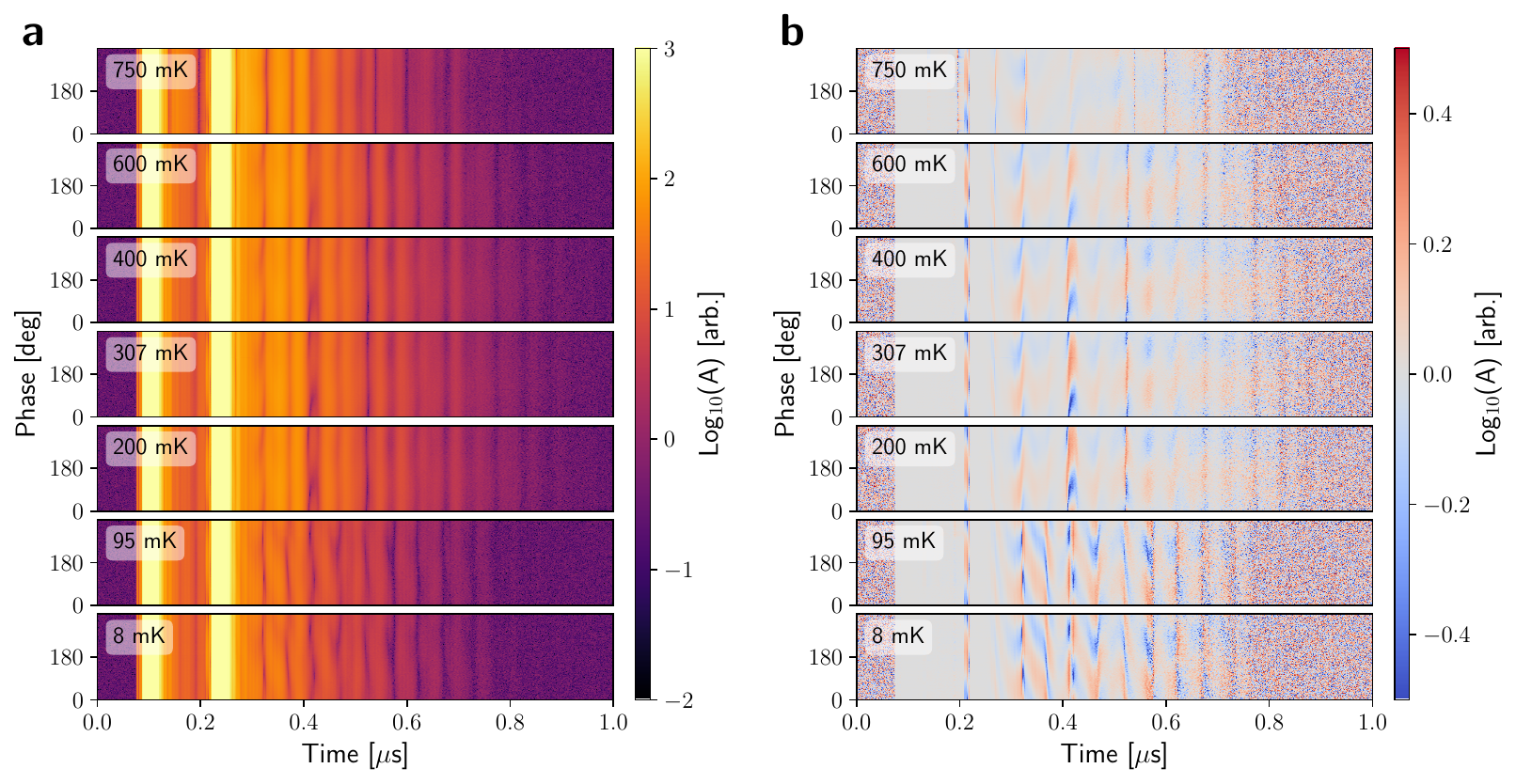}
\caption{Coherent phase control at different temperatures. We perform two-pulse phase coherent control measurements on sapphire samples with spin-coated Shipley 1813 photoresist.  We drive and readout at 3.4 GHz and fix the inter-pulse spacing at 100 ns. The MXC heater is adjusted to set the sample temperature to 8, 95, 200, 307, 400, 600, and 750 mK. At each set point, the system is allowed to thermalize for $>$1 hour before data acquisition. \textbf{a,} Logarithmic amplitude spectrum at different temperatures, showing two control pulses and the resulting transient response. \textbf{b,} Difference plot of (\textbf{a}). At each temperature, the mean transient trace (averaged over all phases) is subtracted from each phase slice. The resulting difference map highlights how the transient signal changes with phase, providing a direct measure of the degree of coherent control. Reduced contrast at higher temperatures indicates weakened phase sensitivity and a loss of coherence.   The apparent increase in ringdown lifetimes between 95 and 200 mK may be attributed to spectral diffusion events, which can shift either the resonance frequency or the coupling strength of the TLSs probed at that frequency. This behavior is then followed by a gradual reduction of the coherent response, which manifests as a monotonic decrease in ringdown lifetimes with increasing temperature.}
\label{fig:phase_control_diff_temp}
\end{figure*}

In this section, we explore coherent control and memory effects at elevated temperatures. We adjust the MXC heater of the dilution refrigerator and vary the temperature from 8 mK to 750 mK, with the system allowed to stabilize for at least one hour at each setpoint to ensure thermalization. We perform phase coherent control measurements on sapphire samples with spin-coated Shipley 1813 photoresist. We send the two pulses at 3.4 GHz and sweep the phase of the first pulse, keeping the phase of the second pulse and the inter-pulse spacing constant at 100 ns. Figure\,\ref{fig:phase_control_diff_temp}\textbf{a} shows the transient spectra at various temperatures. At low temperatures, clear interference fringes appear, indicating coherent control of the transient response and memory retention from the first excitation pulse. As the temperature increases, a pronounced initial change is visible (when comparing the 95 mK and 200 mK spectra), where the enhanced ringdown lifetime may result from a spectral diffusion event that shifted either the resonance frequency or the coupling strength of the TLS response probed at 3.4 GHz. Eventually, these fringes progressively disappear, indicating that the system loses phase sensitivity, while the extracted lifetimes decrease monotonically. To better visualize this evolution, Fig. \ref{fig:phase_control_diff_temp}\textbf{b} shows the differential maps, at each temperature, obtained by subtracting the mean transient trace (averaged over all phases) from each horizontal phase slice. This procedure isolates the phase-dependent oscillatory features, which are clearly visible at low temperatures but gradually fade and reduce as the temperature rises, reinforcing the observation that coherence and memory effects in the TLS ensemble diminish as thermal excitations become dominant.

\section{Effect of Full Thermal Cycling}
\label{app_sec:thermal_cycle}
\begin{figure*}[hpbt!]
\includegraphics[width=\textwidth, keepaspectratio]{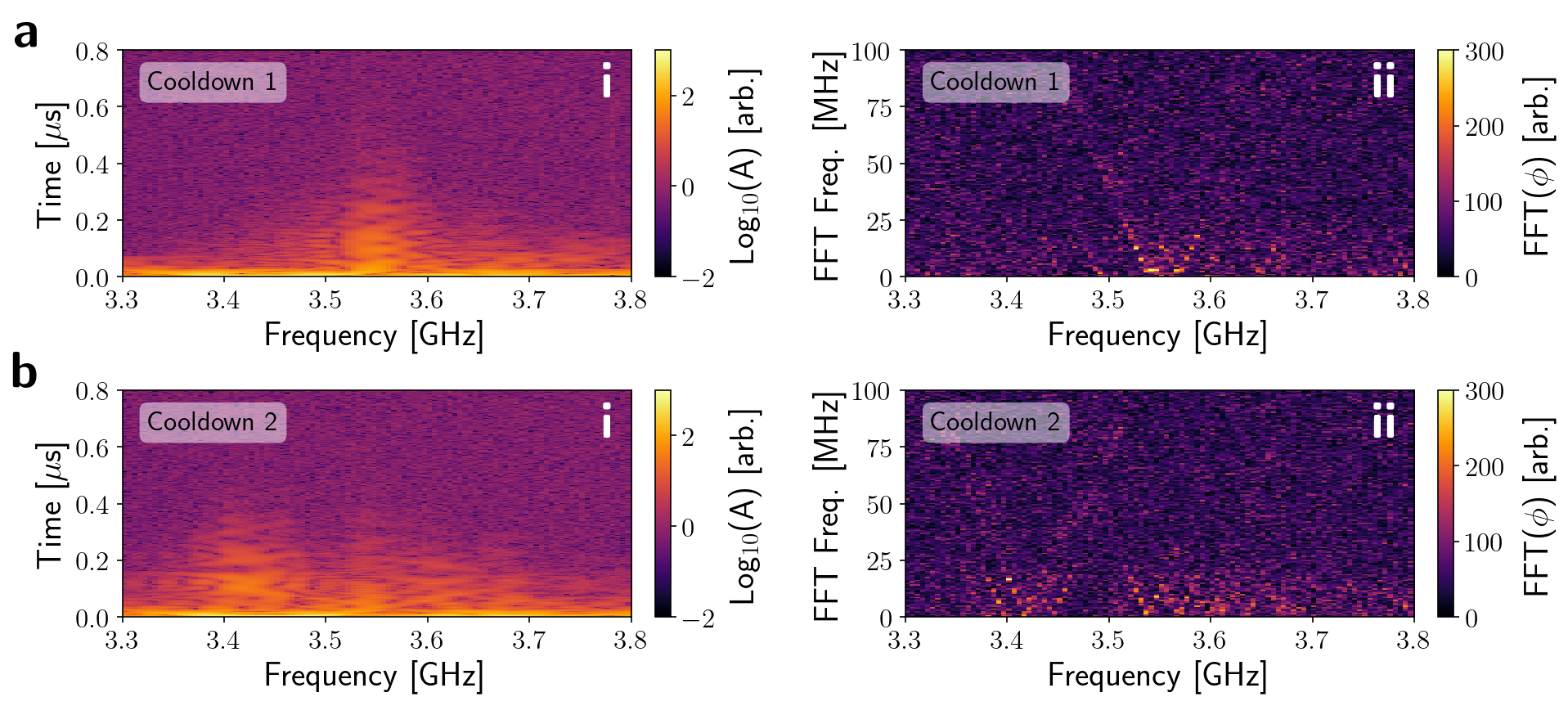}
\caption{BCTDS and phase V analysis before and after thermal cycling. We measure sapphire samples coated with Shipley 1813 photoresist across two separate cooldowns. Both measurements were taken at $\sim$22 mK. We use lower sweep resolution for fast data acquisition. \textbf{a}, Logarithmic amplitude spectrum (\textbf{i}) and FFT of the phase (\textbf{ii}) measured during cooldown 1. \textbf{b}, Corresponding measurements during cooldown 2, following a thermal cycle to room temperature. The two datasets were collected 3 days apart without breaking the dilution fridge vacuum or making any changes to the measurement setup. The frequency span is restricted to 3.3-3.8 GHz to highlight fine structures, which show pronounced differences in both the transient logarithmic amplitude spectra and the phase V features, due to repopulation of TLS defects after thermal cycling.}
\label{fig:thermal_cycle}
\end{figure*}

In Fig.\,\ref{fig:thermal_cycle}, we demonstrate the effect of thermal cycling to room temperature with two back-to-back BCTDS measurements. A sapphire sample coated with Shipley 1813 photoresist is first measured at $\sim$22 mK, then warmed to room temperature, and subsequently cooled back down to $\sim$22 mK without any changes to the experimental setup. We compare the transient logarithmic amplitude spectra and phase FFT spectra over a narrow frequency window (3.3-3.8 GHz) and observe pronounced differences: the ring-down beating patterns change, and the positions of the phase V-structures shift. These modifications are most likely due to morphological changes in the sample that result in reorganization of the local environment around the TLS defects after thermal cycling, a phenomenon widely reported in previous studies \cite{PhysRevLett.105.177001}.

\section{BCTDS Response of Silicon Sample While Sweeping Pulse Amplitude}
\label{app_sec:amp_sweep}

\begin{figure*}[hpbt!]
  \centering 
\includegraphics[width=\textwidth,keepaspectratio]{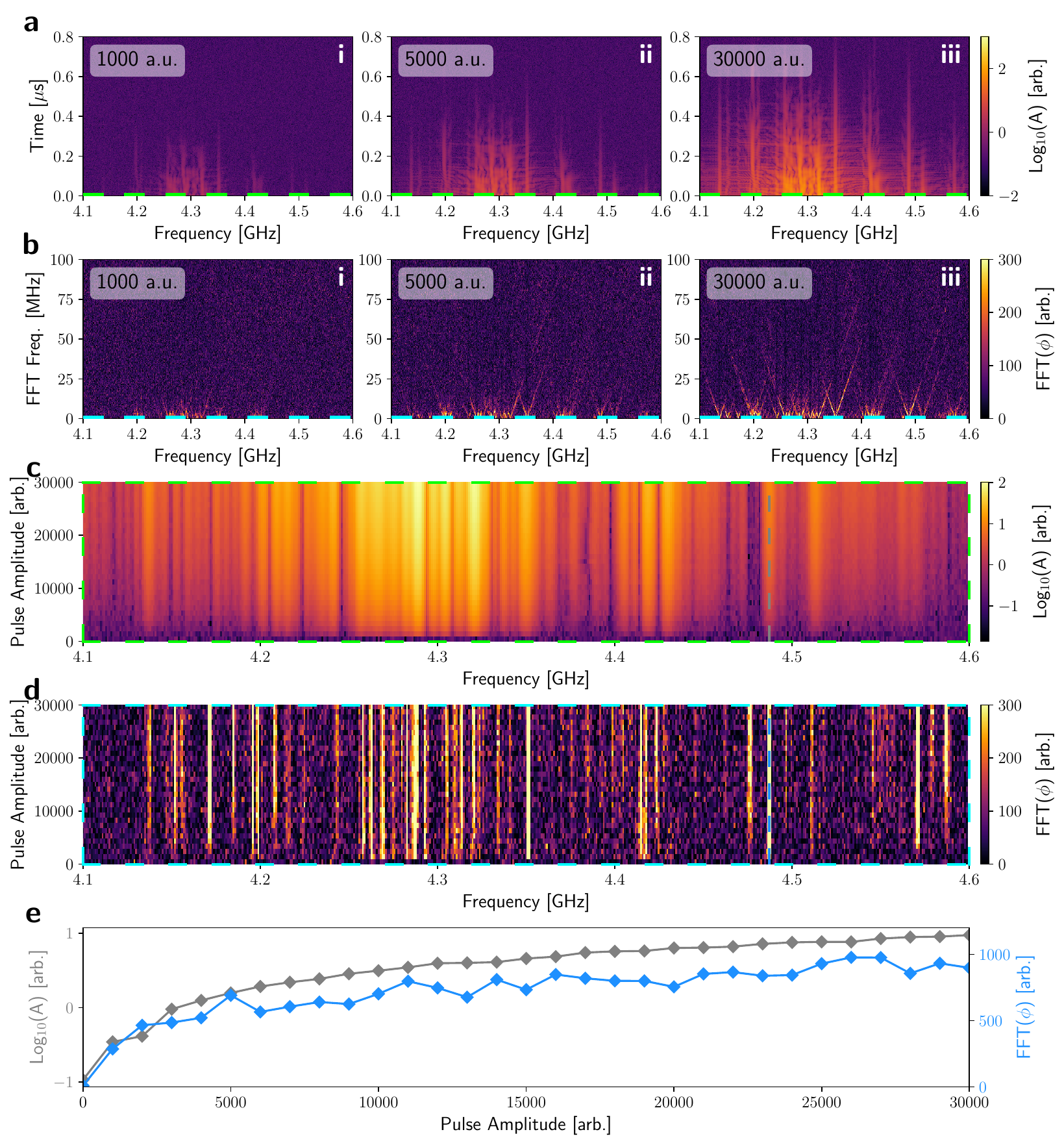}
\caption{BCTDS result for a silicon sample with silicon oxide under different drive amplitudes. The pulse duration is set to be 200 ns. \textbf{a,} Logarithmic amplitude of the homodyne signal arising from the transient dielectric response. \textbf{b,} FFT of the phase of the transient dielectric response, where V-shaped structures centered at the bare eigenfrequencies of TLS defect ensembles are clearly visible. Panels \textbf{a,b} are shown for three different pulse amplitudes: 1000 a.u. (\textbf{i}), 5000 a.u. (\textbf{ii}), and 30000 a.u. (\textbf{iii}). The drive amplitudes across the span are calibrated using a method described in \cite{BCTDS} Appendix E, to ensure even driving across all frequencies. \textbf{c,} Zero-time transient response at different pulse amplitudes. Horizontal slices (lime dashed lines) from the logarithmic amplitude plots (\textbf{a}) are taken at $t = t_{\text{off}} =0$ for a continuous sweep of pulse amplitudes from 0 to 30000 a.u., showing gradual growth of interference patterns as the pulse amplitude increases. \textbf{d,} Base location of phase Vs at different pulse amplitudes. Horizontal slices at 0 MHz (cyan dashed lines) from the phase FFT plots (\textbf{b}) are taken for the 0 to 30000 a.u amplitude sweep. Bright lines marking the eigenfrequencies of the driven system start to emerge, and their positions remain constant as the pulse amplitude increases. \textbf{e,} Vertical cuts of subplots \textbf{c} (gray) and \textbf{d} (blue) at 4.487 GHz (each marked by a black dashed line), illustrating saturation as the pulse amplitude increases.}
\label{fig:amp_sweep}
\end{figure*}

In Fig.\,\ref{fig:phase_V}, we examine the transient response of a silicon sample and the corresponding V-shaped structures in the FFT of phase under varying drive durations. Here, we extend the analysis by exploring their behavior as a function of drive amplitude. As the amplitude increases, interference patterns at later times become more pronounced, as shown in the logarithmic amplitude spectra of Fig.\,\ref{fig:amp_sweep}\textbf{ai–iii}, where we display representative cases at amplitudes of 100, 5000, and 30000 a.u. To capture the full relationship, we take horizontal slices of the logarithmic amplitude spectra at the onset of the transient ring-down and compile them across a continuous sweep of pulse amplitudes from 0 to 30000 arbitrary units (a.u.). The resulting color plot in Fig.\,\ref{fig:amp_sweep}\textbf{c} shows an increase in spectral brightness as the pulse becomes stronger, with minimal shifts in the interference pattern. Phase FFT plots are displayed in Fig.\,\ref{fig:amp_sweep}\textbf{bi–iii}, with zero-frequency slices stacked across amplitudes in Fig.\,\ref{fig:amp_sweep}\textbf{d}. These reveal brighter V-shaped structures as the drive increases, but no changes in their base positions.
Taken together with Fig.\,\ref{fig:phase_V}, these results establish that the locations of the V-bases are independent of both pulse duration and amplitude, supporting their interpretation as the bare eigenfrequencies of the driven system. Finally, the amplitude-sweep data reveal clear saturation behavior. In Fig.\,\ref{fig:amp_sweep}\textbf{e}, we plot vertical slices from Fig.\,\ref{fig:amp_sweep}\textbf{c} (gray) and Fig.\,\ref{fig:amp_sweep}\textbf{d} (blue) at a representative frequency of 4.487 GHz. The response exhibits a rapid increase at low drive, followed by a gradual plateau, consistent with saturation dynamics analogous to acoustic spectral hole burning \cite{Andersson_2021}.

\begin{figure*}[htpb!]
  \centering 
\includegraphics[width=\textwidth,keepaspectratio]{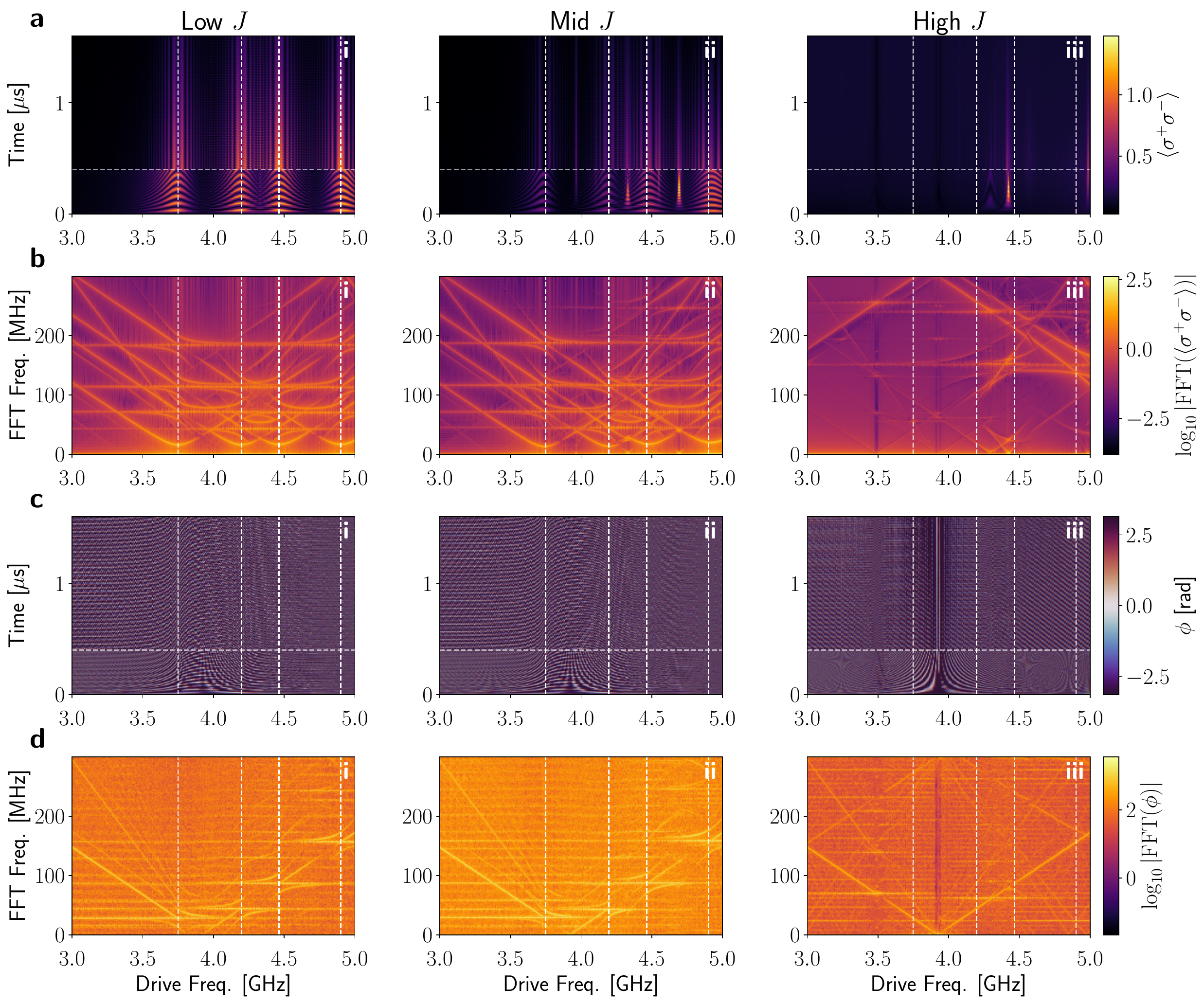}
\caption{Role of dipole–dipole coupling strength. \textbf{a,} Population $\langle\sigma^{+}\sigma^{-}\rangle$ dynamics for an ensemble of $N = 4$ TLS defects driven by a single square pulse of duration $\tau = 400$ \,ns and amplitude $A/2\pi = 100$ \,MHz. The interaction strength is set to $J/2\pi = 5$ MHz  (\textbf{i}), $J/2\pi = 50$ MHz (\textbf{ii}), and $J/2\pi = 0.5$ GHz (\textbf{iii}), with the same labels used in panels~\textbf{b}–\textbf{d}. Bare transition frequencies are drawn from a uniform distribution $\omega_{i}/2\pi\in[3.0,5.0]$\,GHz, and the collective decay rate is $\Gamma/2\pi = 1$\,MHz. Vertical dashed lines in all panels mark the bare frequencies, while the horizontal dashed lines indicate the pulse window in \textbf{a} and \textbf{c}\textbf{b,} Fourier spectra of the population traces in \textbf{a}, showing V-shaped arcs centered on the bare resonance frequencies. Horizontal ridges correspond to the detunings between individual emitters. \textbf{c,} Relative phase between the output-field quadratures, calculated directly from the time-domain data in \textbf{a}. \textbf{d,} Fourier spectra of the phase evolution in \textbf{c}. V-shaped patterns reappear around the bare frequencies, and additional sidebands reveal hybridization and interference, qualitatively consistent with experimental observations.}
\label{fig:role_of_interactions}
\end{figure*}

\section{Floquet Derivation} \label{app_sec:floquet}
We first briefly summarize the formalism following \cite{Giovannini_2020, PhysRevB.68.165315}. 
Consider a time-periodic Hamiltonian \(H(t)\) with period \(T\), $H(t + T) = H(t)$ with Floquet frequency $\Omega = 2\pi / T$. Floquet theorem states that solutions to the time-dependent Schrödinger equation can be expressed as
\begin{equation}
|\psi_\alpha(t)\rangle = e^{-i E_\alpha t} |u_\alpha(t)\rangle, \quad |u_\alpha(t+T)\rangle = |u_\alpha(t)\rangle,
\end{equation}
where the Floquet mode \( |u_\alpha(t)\rangle \) is periodic, and \(E_\alpha\) is the quasi-energy. Expanding \( |u_\alpha(t)\rangle \) into Fourier harmonics,
\begin{equation}
|u_\alpha(t)\rangle = \sum_{m=-\infty}^\infty e^{-i m \Omega t} |u_\alpha^m\rangle,
\end{equation}
leads to the eigenvalue equation in the extended space:
\begin{equation}
\sum_m \mathcal{H}_{nm} |u_\alpha^m\rangle = E_\alpha |u_\alpha^n\rangle, \quad \mathcal{H}_{nm} = H_{n-m} + m \Omega \delta_{nm}
\end{equation}
where
\begin{equation}
\quad H_k = \frac{1}{T} \int_0^T dt\, e^{i k \Omega t} H(t)
\end{equation}
is the $k$-th Fourier component of the $T$-periodic Hamiltonian $H(t)$. For a monochromatic drive,
\begin{equation}
H(t) = H_0 + V e^{i \Omega t} + V^\dagger e^{-i \Omega t},
\end{equation}
the Floquet Hamiltonian \(\mathcal{H}\) is block-tridiagonal
\begin{equation}
\mathcal{H} = 
\begin{pmatrix}
\ddots & \vdots & \vdots & \vdots & \iddots \\
\cdots & H_0 - \Omega & V & 0 & \cdots \\
\cdots & V^\dagger & H_0 & V & \cdots \\
\cdots & 0 & V^\dagger & H_0 + \Omega & \cdots \\
\iddots & \vdots & \vdots & \vdots & \ddots
\end{pmatrix}.    
\end{equation}
For a dipole operator $\hat{P}(t)$, the dipole response function is given by the retarded commutator
\begin{equation}
\chi(t,t') = -i \theta(t - t') \langle [\hat{P}(t), \hat{P}(t')] \rangle,
\end{equation}
where \(\theta\) is the Heaviside step function and \(\langle \cdot \rangle\) is the expectation in the steady-state Floquet. 
Using Floquet states, the response can be expanded as
\begin{equation}
\begin{aligned}
\chi(t,t') &= \sum_{n,m} e^{-i n \Omega t} e^{i m \Omega t'} \chi_{nm}(t - t')\\
& =\sum_{n,m} e^{-i n \Omega t} e^{i m \Omega t'} \int \frac{d \omega}{2 \pi} e^{-i\omega (t - t')} \chi_{nm}(\omega) 
\end{aligned}
\end{equation}
where the Fourier components satisfy the Dyson equation
\begin{equation}
\chi_{nm}(\omega) = \chi_{nm}^0(\omega) + \sum_{k} \chi_{nk}^0(\omega) V_{km} \chi_{km}(\omega).
\end{equation}
Explicitly, the Fourier transform of the dipole response function is expressed via the Floquet Green's functions as
\begin{equation}
\chi_{nm}(\omega) = \sum_{\alpha,\beta} \sum_{\ell \in \mathbb{Z}} \frac{d_{\alpha \beta}^\ell d_{\beta \alpha}^{m - n - \ell}}{\omega - (E_\alpha - E_\beta + \ell \Omega) + i \eta},
\end{equation}
where
\[
d_{\alpha \beta}^m = \frac{1}{T} \int_0^T dt\, e^{i m \Omega t} \langle u_\alpha(t) | \hat{P} | u_\beta(t) \rangle,
\]
are the elements of the Floquet dipole matrix, \(\eta \to 0^+\). The poles occur at
\begin{equation}
\omega = E_\alpha - E_\beta + \ell \Omega,
\end{equation}
which determines the resonance frequencies.

When the system is driven by a periodic square pulse of duration \(\tau\), then turned off at \(t = \tau\), the system freely evolves with no drive for \(t > \tau\). The postpulse time evolution contains oscillatory terms at frequencies set by quasi-energy differences \(E_\alpha - E_\beta\) and sideband shifts \(\ell \Omega\), weighted by the residues of these poles in \(\chi\).

\subsection{Role of Interactions}

We explore the role of TLS–TLS interactions in shaping the quantum dynamics of the system. The main objective of this section is to examine how the spectral features of the phase evolution (referred to as phase-Vs) change with interaction strength. Quantitative estimates of dipole–dipole interactions between TLS defects have been reported by Lisenfeld \emph{et al.}~\cite{Lisenfeld2015}, with transverse and longitudinal components on the order of $\sim150$~MHz and $\sim400$~MHz, respectively. This corresponds to a net coupling strength of about 800~MHz, which remains smaller than the individual TLS frequencies. In our case, however, we drive the entire TLS ensemble over a broad frequency range. When TLSs are close in frequency, resonant exchange interactions are expected to dominate,  leading to coherent swapping of excitations between them. This resonant exchange gives rise to hybridized states and oscillatory dynamics characteristic of coherent coupling. However, when the TLS are detuned or experience strong decoherence, the flip–flop processes becomes energetically suppressed, and the interaction reduces to a static, dispersive (Ising-like) coupling that merely shifts their energy levels, leading to spectral diffusion of the entire ensemble. 

Figure \ref{fig:role_of_interactions}\textbf{a} shows the dynamical response of a TLS ensemble subject to a finite-duration square-envelope drive with carrier $\cos(\omega_d t)$, highlighting how dipole–dipole interactions modify the system behavior. In the weakly interacting limit, TLS defects evolve independently, giving rise to sharp resonance features in $\langle \sigma^+ \sigma^- \rangle$ corresponding to coherent Rabi oscillations near their transition frequencies. The finite pulse duration broadens the response with a sinc-like frequency profile, partially exciting off-resonant TLS. As the interaction strength increases, TLS become coherently coupled and evolve collectively. This is reflected in the Fourier spectrum of $\langle \sigma^+ \sigma^- \rangle$ in Fig.\ref{fig:role_of_interactions}\textbf{b}, where vertical bands from individual TLS deform into avoided crossings, a signature of hybridization and level repulsion. The splitting at these crossings encodes the dipole–dipole coupling strength, while in the strong-interaction regime the drive couples predominantly to bright collective modes, leaving others dark due to destructive interference.

\begin{figure}[htpb!]
\begin{centering}
\includegraphics[width=0.5\textwidth]{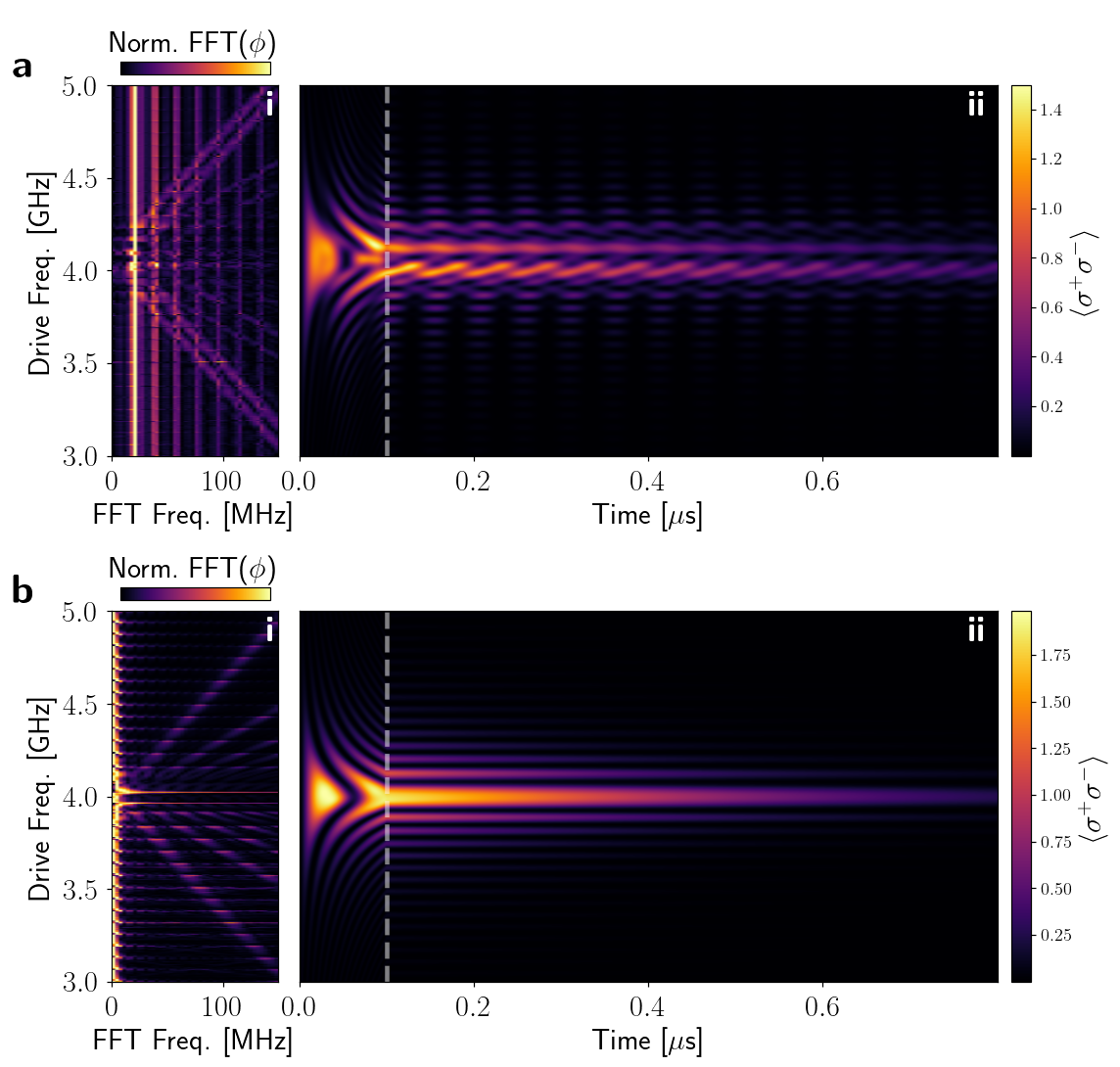}
\end{centering}
\caption{Effect of detuning on the dynamics and spectral response of an interacting TLS ensemble with $N=2$. \textbf{a}, Slightly off-resonant TLSs with resonance frequencies $\omega_1/2\pi = 4.0$ GHz and $\omega_2/2\pi = 4.12$ GHz. \textbf{b}, Degenerate TLSs with $\omega_1/2\pi = \omega_2/2\pi = 4.0$ GHz. We set the dipole–dipole coupling strength to $J/2\pi = 10$ MHz, the drive amplitude to $A/2\pi = 100$ MHz, the pulse duration to $\tau = 100$ ns (denoted by the vertical dashed lines in \textbf{aii} and \textbf{bii}), and the collective dissipation rate to $\Gamma/2\pi = 2$ MHz. \textbf{i} Fast Fourier transform (FFT) of the phase evolution. \textbf{ii} Population dynamics of the TLS ensemble. For the off-resonant case, the dynamics exhibit beating between the nearby resonances and corresponding sidebands. In the degenerate case, a single resonance appears without beating. In both cases, the excitation decays according to the collective radiative dissipation of the system.}
\label{fig:braiding_physics}
\end{figure}

The phase response, $\phi = \arg(\langle \sigma^+ \rangle)$, provides complementary information, as shown in Fig.\ref{fig:role_of_interactions}\textbf{c}. For weak interactions, $\phi$ evolves smoothly with detuning, but stronger coupling produces structured maps with rapid gradients, sheared fringes, and abrupt $\pi$ or $2\pi$ jumps, particularly near avoided crossings. These features arise from interference between nearly degenerate collective modes, where small energy splittings cause slow relative phase accumulation, a manifestation of dephasing-induced critical slowing down. While amplitude is suppressed in these regions, the phase remains highly structured, indicating persistent coherence through interaction-driven synchronization. This is corroborated in the Fourier spectrum of $\phi$ in Fig.\ref{fig:role_of_interactions}\textbf{d}, which exhibits narrowband features even when the amplitude spectrum appears diffuse, serving as a fingerprint of interaction-induced synchronization. For weak interactions, distinct avoided crossings appear near the bare TLS frequencies, indicated by the white dashed lines in Fig.~\ref{fig:role_of_interactions}\textbf{d}. As the interaction strength increases, a sharp spectral feature emerges while the avoided crossings, or level repulsion, observed in the phase evolution become substantially suppressed, resulting in a single apparent V-shaped pattern that suggests collective spin-like behavior. This observation may indicate that TLSs close in frequency form clusters characterized by strong intra-cluster transverse interactions ($\hat{\sigma}_x\hat{\sigma}_x$) and weaker inter-cluster longitudinal couplings ($\hat{\sigma}_z\hat{\sigma}_z$), which promote spectral diffusion and shift the effective eigenfrequencies of the collective ensemble.

Scaling the driven interacting few-spin system to larger ensembles is challenging due to the exponential growth of the Hilbert space with system size. Several approaches may mitigate this cost while capturing the essential physics. When the TLSs are nearly identical and evolve symmetrically, collective-spin (Dicke) mappings reduce the dimensionality from $2^N$ to $N+1$, efficiently describing cooperative effects. For systems with moderate disorder or short-range correlations, cluster mean-field or cumulant-expansion methods provide a controlled approximation of local correlations \cite{Kraemer2015}. Tensor-network techniques, such as matrix product states or operators with time-evolving block decimation evolution  (TEBD) \cite{TEBD2016, TEBD2020}, compress the many-body state efficiently in low-entanglement regimes and can include long-range couplings through suitable approximations. Dissipative dynamics can be treated using quantum-trajectory (Monte Carlo wavefunction) methods, replacing the full density-matrix evolution with averages over pure-state trajectories that parallelize naturally. In the weak-excitation limit, Holstein–Primakoff or bosonic truncations may further reduce the computational load. An alternative approach incorporates the full electromagnetic environment by solving the Maxwell--Bloch equations in a realistic waveguide geometry using finite-difference time-domain (FDTD) simulations \cite{Zhou2024_FDTD_TLS}. This method naturally accounts for spatial structure, retardation, boundary effects, and field-mediated interactions, allowing large TLS ensembles to be modeled without explicitly constructing the full $2^N$-dimensional Hilbert space. Together, these methods provide complementary routes to extend the TLS model to experimentally relevant system sizes while retaining the key quantum and collective effects.

In summary, our numerical simulations reveal avoided crossings and structured phase dynamics that highlight the interplay between external driving and many-body interactions. The drive imposes coherence over a broad spectral range, while interactions select and stabilize specific collective modes. As the coupling strength increases, the ensemble transitions from independent TLS dynamics to a regime dominated by synchronized phase evolution and selective mode excitation, marking the emergence of sharp spectral features that may indicate clustering of strongly interacting TLS defects, and such a cluster may behave as an effective spin giving rise to a single V in the spectrum. The detailed quantitative characterization of spectral diffusion, broad distribution of coupling strengths, as well as inter and intra-cluster interactions dominated by longitudinal and transverse dipole-dipole interactions, as well as the effect of thermal cycling on the TLS ensemble dynamics will be addressed in future work.

\subsection{Role of Detuning}
\label{app_sec:role_of_detuning}

We now examine the role of detuning in shaping both the system dynamics and the spectral features observed experimentally. Fig.\,\ref{fig:braiding_physics} presents the (\textbf{ii}) population dynamics and (\textbf{i}) FFT of the phase for a system of $N=2$ TLSs under two conditions: (\textbf{a}) slightly off-resonant TLSs, and (\textbf{b}) degenerate TLSs. In Fig.\ref{fig:braiding_physics}\textbf{a, ii}, the population dynamics exhibit beating, arising from the interference of nearby resonances, with decay determined by the collective radiative dissipation. The corresponding FFT (Fig.\ref{fig:braiding_physics}\textbf{a, i}) shows two shifted V-shaped features converging towards the individual TLS frequencies, together with dressed sidebands. In contrast, the degenerate case (Fig.\ref{fig:braiding_physics}\textbf{b}) displays a smooth ring-down in the population without beating. The FFT in this case exhibits a single, centered V-shaped structure at the common TLS frequency, accompanied by a strong DC component. Notably, similar features appear in the experimental spectra of Fig.\ref{fig:phase_V}\textbf{a,c,iii}, around 4.2 and 4.5~GHz. These results indicate that the broadband nature of our spectroscopy allows us to probe multiple TLS configurations, capturing both slightly detuned interactions and effectively degenerate cases in different parts of the spectrum. As a result, the ensemble dynamics reveal a rich variety of spectral features that provide insight into TLS defects and support the numerical model presented here.


\end{document}